\documentclass[twocolumn,twocolappendix,times]{aastex631}
\usepackage[utf8]{inputenc}
\usepackage{fp}
\usepackage{hyperref}

\received{2021 September 20}
\revised{2022 March 31}
\accepted{2022 April 5}
\submitjournal{\textit{The Astrophysical Journal}}

\begin{document}
\FPset\gold{15}
\FPset\silver{11}
\FPset\control{8}
\FPeval\bumps{clip(\gold+\silver)}
\FPeval\total{clip(\bumps+\control)}
\FPeval\goldpct{round(\gold/\total*100,0)}
\FPeval\silverpct{round(\silver/\total*100,0)}
\FPeval\controlpct{round(\control/\total*100,0)}
\FPeval\bumpspct{round(\bumps/\total*100,0)}

\title{\large Bumpy Declining Light Curves Are Common in Hydrogen-poor Superluminous Supernovae}
\shorttitle{Bumpy Declining Light Curves Are Common in SLSNe}
\shortauthors{Hosseinzadeh et al.}

\correspondingauthor{Griffin Hosseinzadeh}
\email{griffin0@arizona.edu}

\newcommand{\UA}{\affiliation{Steward Observatory, University of Arizona, 933 North Cherry Avenue, Tucson, AZ 85721-0065, USA}}
\newcommand{\CfA}{\affiliation{Center for Astrophysics \textbar{} Harvard \& Smithsonian, 60 Garden Street, Cambridge, MA 02138-1516, USA}}
\newcommand{\Columbia}{\affiliation{Department of Physics and Columbia Astrophysics Laboratory, Columbia University, Pupin Hall, New York, NY 10027, USA}}
\newcommand{\Flatiron}{\affiliation{Center for Computational Astrophysics, Flatiron Institute, 162 5th Avenue, New York, NY 10010-5902, USA}}
\newcommand{\STSci}{\affiliation{Space Telescope Science Institute, 3700 San Martin Drive, Baltimore, MD 21218, USA}}
\newcommand{\Birmingham}{\affiliation{Birmingham Institute for Gravitational Wave Astronomy and School of Physics and Astronomy, University of Birmingham, Birmingham B15 2TT, UK}}
\newcommand{\CIERA}{\affiliation{Center for Interdisciplinary Exploration and Research in Astrophysics and Department of Physics and Astronomy, \\Northwestern University, 1800 Sherman Avenue, 8th Floor, Evanston, IL 60201, USA}}

\author[0000-0002-0832-2974]{Griffin Hosseinzadeh}
\UA\CfA

\author[0000-0002-9392-9681]{Edo~Berger}
\CfA

\author[0000-0002-4670-7509]{Brian~D.~Metzger}
\Columbia\Flatiron

\author[0000-0001-6395-6702]{Sebastian~Gomez}
\STSci\CfA

\author[0000-0002-2555-3192]{Matt~Nicholl}
\Birmingham

\author[0000-0003-0526-2248]{Peter~Blanchard}
\CIERA

\begin{abstract}
Recent work has revealed that the light curves of hydrogen-poor (Type~I) superluminous supernovae (SLSNe), thought to be powered by magnetar central engines, do not always follow the smooth decline predicted by a simple magnetar spin-down model. Here we present the first systematic study of the prevalence and properties of ``bumps'' in the post-peak light curves of 34 SLSNe. We find that the majority (44--76\%) of events cannot be explained by a smooth magnetar model alone. We do not find any difference in supernova properties between events with and without bumps. By fitting a simple Gaussian model to the light-curve residuals, we characterize each bump with an amplitude, temperature, phase, and duration. We find that most bumps correspond with an increase in the photospheric temperature of the ejecta, although we do not see drastic changes in spectroscopic features during the bump. We also find a moderate correlation ($\rho\approx0.5$; $p\approx0.01$) between the phase of the bumps and the rise time, implying that such bumps tend to happen at a certain ``evolutionary phase,'' $(3.7\pm1.4)t_\mathrm{rise}$. Most bumps are consistent with having diffused from a central source of variable luminosity, although sources further out in the ejecta are not excluded. With this evidence, we explore whether the cause of these bumps is intrinsic to the supernova (e.g., a variable central engine) or extrinsic (e.g., circumstellar interaction). Both cases are plausible, requiring low-level variability in the magnetar input luminosity, small decreases in the ejecta opacity, or a thin circumstellar shell or disk.
\end{abstract}


\section{Introduction\label{sec:intro}}
In the decade since their discovery \citep{chomiuk_pan-starrs1_2011,quimby_hydrogen-poor_2011,gal-yam_luminous_2012}, a major focus in the study of hydrogen-poor (Type~I) superluminous supernovae (SLSNe) has been to identify the power source(s) behind their extreme and long-lived luminosities. Four major mechanisms have been proposed, alone or in combination: central energy injection by a newly born magnetar \citep{kasen_supernova_2010,woosley_bright_2010}, central energy injection from fallback accretion onto a newly born black hole \citep{dexter_supernova_2013}, radioactive decay of an extremely large mass of $^{56}$Ni (as expected for a pair-instability supernova (SN); \citealt{kasen_pair_2011,dessart_superluminous_2012}), and/or kinetic energy released through collision with circumstellar material (CSM; as expected for a pulsational pair-instability SN; \citealt{woosley_pulsational_2007}). Each of these proposals has strengths and weaknesses, but in general, the magnetar model has had the most success in explaining Type~I SLSN light curves and spectra (e.g., \citealt{nicholl_magnetar_2017}; see \citealt{gal-yam_most_2019} for a review). Combinations of these models are also possible, e.g., accretion onto a central magnetar \citep{metzger_effects_2018} or interaction between magnetar-powered ejecta and CSM \citep{chatzopoulos_extreme_2016,li_energy_2020}, but are difficult to constrain with the limited observational data available.

Simple magnetic dipole theory predicts a smooth spin-down process ($L \propto t^{-2}$; \citealt{ostriker_pulsars_1971}). If magnetars do spin down following the simple dipole formula, and this is the only power source heating the ejecta, we would expect the light curves of magnetar-powered SNe to be equally smooth. However, several authors have noticed ``bumps'' in SLSN light curves, both early bumps before the main peak (e.g., \citealt{leloudas_sn_2012,nicholl_diversity_2015,smith_des14x3taz:_2016,angus_superluminous_2019}) and late bumps or ``undulations'' in the declining light curve (e.g., \citealt{inserra_super-luminous_2013,nicholl_superluminous_2014,inserra_complexity_2017,nicholl_sn_2016,fiore_sn_2021}). Pre- and post-maximum bumps likely have distinct physical origins. \cite{piro_using_2015} and \cite{kasen_magnetar-driven_2016} explain the former as shock breakout through preexisting CSM and magnetar-inflated ejecta, respectively, whereas \cite{margalit_grb-slsn_2018} suggest the breakout of a relativistic jet. \cite{moriya_dip_2012} argue that pre-peak bumps are in fact dips caused by changes in CSM opacity. Post-peak bumps could result from collisions between the ejecta and multiple shells or clumps of CSM, as predicted in pulsational pair-instability SNe \citep{woosley_pulsational_2007} or the explosions of rapidly rotating Wolf--Rayet stars \citep{aguilera-dena_related_2018}. Alternatively, they could occur when the magnetar engine flares \citep{yu_possible_2017} and/or drives ionization fronts through the ejecta \citep{metzger_ionization_2014}. Recently, \cite{vurm_gamma-ray_2021} modeled in detail the thermalization process for the $\gamma$-rays produced by a magnetar central engine, showing that more complex light curves are expected, even without an additional power source. However, even their detailed model does not reproduce the shorter bumps in the light curves of SNe~2015bn and 2017egm \citep{nicholl_sn_2016,nicholl_superluminous_2017,bose_gaia17biu/sn_2018}.

Post-peak bumps are thus far poorly understood and characterized for only a small number of events, due to the difficulty of observing distant, fading SLSNe for hundreds of days. In the past, most SLSN light curves were not sampled frequently enough to observe (or rule out) such variability. However, today's time-domain surveys are beginning to accumulate large samples of well-observed events (e.g., \citealt{de_cia_light_2018}, \citealt{lunnan_hydrogen-poor_2018}, \citealt{angus_superluminous_2019}). The purpose of this work is to systematically investigate the prevalence and characteristics of bumps in the declining light curves of Type~I SLSNe, with a view toward constraining their origin.

We first assemble a sample of SLSNe with well-sampled late-phase light curves and model them with a magnetar spin-down model to determine if there is any excess unexplained emission (\S\ref{sec:model}). After characterizing these excesses, we search for any trends within or between the SN parameters and the bump parameters (\S\ref{sec:analysis}). We then discuss how these clues might favor one of two interpretations for the power source of the bumps: interaction with a CSM shell or changes in the central engine luminosity (\S\ref{sec:interp}).

\section{Modeling\label{sec:model}}
\subsection{Sample Selection\label{sec:sample}}
We begin by assembling a ``master list'' of all transients classified as Type~I SLSNe in the existing literature, including all SLSNe announced on the Transient Name Server or via The Astronomer's Telegram, regardless of whether they have been included in a refereed publication. As far as we are aware, this is the most complete list of SLSNe available, with 206 events as of this writing.\footnote{This includes events with peak absolute magnitudes as faint as $-19$, provided that they have SLSN-like spectra.} Where data are publicly available, we have attempted to verify the SLSN classification through spectral comparisons and/or the absolute magnitude; details will be presented by S.~Gomez et al.\ (2022, in preparation).

From this list, we select SLSNe with light curves that are well sampled months after peak,  when  bumps have been seen in previous work. We do not require that there be a post-peak bump, but rather that a bump would be detectable if present during these phases. We identify \total{} SLSNe with sufficient data, which are listed in Table~\ref{tab:sne}. All the SLSNe in our final sample have at least 12 observations in a single band out to at least 64 rest-frame days after pseudobolometric maximum light. Of the final sample, the only questionable classifications are DES16C3cv and SN~2018fcg, which both have spectra that are redder than expected around maximum light and peak around $M \approx -20$~mag. These do not significantly affect our statistics, and in fact they nearly cancel out in the bump prevalence calculation (\S\ref{sec:prevalence}), so we proceed to include them in our sample.

\begin{deluxetable*}{lcll}
\tablecaption{Final SLSN Sample\label{tab:sne}}
\tablecolumns{4}
\tablehead{\colhead{Name} & \colhead{Redshift} & \colhead{Photometry Reference(s)} & \colhead{MOSFiT Model Reference}}
\startdata
\cutinhead{Gold Sample (\gold{})}
SN 2007bi & 0.128 & \cite{gal-yam_supernova_2009}, \cite{young_two_2010} & \cite{nicholl_magnetar_2017} \\
SN 2010md & 0.099 & \cite{inserra_super-luminous_2013}, \cite{de_cia_light_2018}; SOUSA & \cite{nicholl_magnetar_2017} \\
SN 2011ke & 0.143 & \cite{inserra_super-luminous_2013}, \cite{de_cia_light_2018}; SOUSA & \cite{nicholl_magnetar_2017} \\
PS1-12cil\textsuperscript{a} & 0.32\phn & \cite{lunnan_hydrogen-poor_2018} & this work \\
SN 2015bn\textsuperscript{a} & 0.114 & \cite{nicholl_sn_2016,nicholl_superluminous_2016,nicholl_one_2018}, CPCS (Alert 25794), CSS, SOUSA & \cite{nicholl_magnetar_2017} \\
DES16C3cv & 0.727 & \cite{angus_superluminous_2019} & \cite{hsu_magnetar_2021} \\
SN 2016ard & 0.203 & \cite{blanchard_type_2018} & \cite{blanchard_type_2018} \\
SN 2017egm\textsuperscript{a} & 0.031 & \cite{nicholl_superluminous_2017}, \cite{bose_gaia17biu/sn_2018}, Gaia, PS1, SOUSA, this work & this work \\
SN 2017gci\textsuperscript{a} & 0.087 & \cite{fiore_sn_2021} & this work \\
SN 2018bym & 0.274 & \cite{lunnan_four_2020}, ATLAS, CSS, PS1, this work & this work \\
SN 2018kyt\textsuperscript{b} & 0.108 & \cite{yan_helium-rich_2020}, Gaia, ZTF & this work \\
SN 2019hge\textsuperscript{b} & 0.867 & \cite{yan_helium-rich_2020}, ATLAS, Gaia, PS1, ZTF & this work \\
SN 2019neq & 0.108 & ATLAS, PS1, ZTF, this work & this work \\
SN 2019ujb & 0.165 & ATLAS, Gaia, ZTF, this work & this work \\
SN 2019unb\textsuperscript{b} & 0.064 & \cite{yan_helium-rich_2020}, ATLAS, Gaia, PS1, ZTF & this work \\
\cutinhead{Silver Sample (\silver{})}
PS1-11ap & 0.524 & \cite{mccrum_superluminous_2014}, \cite{lunnan_hydrogen-poor_2018} & \cite{nicholl_magnetar_2017} \\
LSQ12dlf & 0.255 & \cite{nicholl_superluminous_2014} & \cite{nicholl_magnetar_2017} \\
PTF12hni & 0.107 & \cite{de_cia_light_2018} & \cite{villar_superluminous_2018} \\
SSS120810 & 0.156 & \cite{nicholl_superluminous_2014}, CSS & \cite{nicholl_magnetar_2017} \\
iPTF13cjq & 0.396 & \cite{de_cia_light_2018} & \cite{villar_superluminous_2018} \\
CSS130912\textsuperscript{c} & 0.431 & \cite{vreeswijk_early-time_2017} & \cite{nicholl_magnetar_2017} \\ 
LSQ14mo & 0.253 & \cite{chen_evolution_2017}, SOUSA & \cite{nicholl_magnetar_2017} \\
LSQ14bdq & 0.345 & \cite{nicholl_diversity_2015} & \cite{nicholl_magnetar_2017} \\
SN 2016wi\textsuperscript{d} & 0.224 & \cite{yan_hydrogen-poor_2017}, PS1 & \cite{nicholl_magnetar_2017} \\ 
SN 2017dwh & 0.13\phn & \cite{blanchard_hydrogen-poor_2019} &\cite{blanchard_hydrogen-poor_2019} \\
SN 2019lsq & 0.14\phn & ATLAS, Gaia, ZTF & this work \\
\cutinhead{Control Sample (\control{})}
PTF09cnd & 0.259 & \cite{quimby_hydrogen-poor_2011}, \cite{de_cia_light_2018}, SOUSA & \cite{nicholl_magnetar_2017} \\
SN 2010gx & 0.230 & \cite{pastorello_ultra-bright_2010}, \cite{quimby_hydrogen-poor_2011}, \cite{de_cia_light_2018}, SOUSA & \cite{nicholl_magnetar_2017} \\
PTF12dam & 0.107 & \cite{nicholl_slowly_2013}, \cite{vreeswijk_early-time_2017}, SOUSA & \cite{nicholl_magnetar_2017} \\
PTF12gty & 0.176 & \cite{de_cia_light_2018} & \cite{villar_superluminous_2018} \\
iPTF13ajg & 0.740 &  \cite{vreeswijk_hydrogen-poor_2014} & \cite{nicholl_magnetar_2017} \\
DES15S2nr & 0.220 & \cite{angus_superluminous_2019} & \cite{hsu_magnetar_2021} \\
iPTF16bad & 0.247 & \cite{yan_hydrogen-poor_2017} & \cite{nicholl_magnetar_2017} \\
SN 2018fcg & 0.101 & ATLAS, Gaia, ZTF, this work & this work \\
\enddata

\tablecomments{See Appendix for details on the new photometry presented in this work. \textit{Public photometry sources}: Asteroid Terrestrial-impact Last Alert System (ATLAS; \citealt{tonry_atlas:_2018}); Cambridge Photometric Calibration Server (CPCS; \url{http://gsaweb.ast.cam.ac.uk/followup/}); Catalina Sky Survey (CSS; \citealt{drake_first_2009}); Gaia Photometric Science Alerts (\url{http://gsaweb.ast.cam.ac.uk/alerts}); Panoramic Survey Telescope and Rapid Response System 1 (PS1; \citealt{chambers_pan-starrs1_2016}); Swift Optical/Ultraviolet Supernova Archive (SOUSA; \citealt{brown_sousa:_2014}); Zwicky Transient Facility (ZTF; \citealt{bellm_zwicky_2019}). \textit{Photometry aggregators and brokers}: Finding Luminous and Exotic Extragalactic Transients \citep{gomez_fleet:_2020}; Make Alerts Really Simple (\url{https://mars.lco.global}); Open Supernova Catalog \citep{guillochon_open_2017}; Transient Name Server (\url{https://wis-tns.org}).}

\textsuperscript{a}{Light curves show two distinct bumps (Section~\ref{sec:model3}).}

\textsuperscript{b}{Light curves show two distinct bumps (Section~\ref{sec:model3}) and spectra show helium lines \citep{yan_helium-rich_2020}.}

\textsuperscript{c}{Previously published under the name iPTF13dcc.}

\textsuperscript{d}{Previously published under the name iPTF15esb.}

\end{deluxetable*}
\vspace{-24pt}]

\subsection{Magnetar Modeling\label{sec:model1}}
We model these light curves using a magnetar-powered SLSN model of \cite{nicholl_magnetar_2017} implemented in the Modular Open-source Fitter for Transients \citep[MOSFiT;][]{guillochon_mosfit:_2017,guillochon_mosfit:_2018}. In many cases, these SNe have already been modeled by previous authors (last column of Table~\ref{tab:sne}), in which case we use their published models except in one of two cases. If the fit is affected by an early bump before the main peak, we exclude any decreasing points before the main peak and refit the light curve. In addition, if the data set includes multiple reductions of the same images, we choose only one reduction and refit the light curve. For more recent SNe, we perform the fitting ourselves with the same software and model used in previous work \citep[see][]{nicholl_magnetar_2017,villar_superluminous_2018,blanchard_pre-explosion_2020,hsu_magnetar_2021}. We run each model until convergence, as measured by a potential scale reduction factor \citep{gelman_inference_1992} of 1.2. The new and refined models are available on Zenodo \citep{hosseinzadeh_magnetar+bump_2022}.

As many previous authors have demonstrated, the magnetar model generally fits these data well. However, we are concerned with the details of when and how the SLSN light curves deviate from the smooth magnetar model. For each SN in our sample, we subtract the best-fit model from our observed light curve and plot the residuals in each band. Figure~\ref{fig:mosfit} (left) shows an example for SN~2011ke. Note in particular the ``bump'' consisting of positive residuals, coherent across all observed bands, that peaks about 50 days after maximum light. Of the \total{} events in our final sample, \gold{} (\goldpct{}\%) show significant deviation from their magnetar models after maximum light, an additional \silver{} (\silverpct{}\%) show possible deviation, and \control{} (\controlpct{}\%) show no deviation. We label these the gold, silver, and control samples, respectively (see Table~\ref{tab:sne}). Bumps are identified via visual inspection, rather than with a strict detection threshold. Thus, the percentages above should be treated only as rough estimates. However, as a point of reference, all of the bumps have at least one point that is ${>}3\sigma$ above the median model, all of the gold bumps have at least one point that is ${>}5\sigma$ above the median model, and most of the bumps have points ${>}10\sigma$ above the median model. Better time sampling, coherence across multiple bands, and larger residuals (further outside the range of possible magnetar models) contribute to the assignment of a ``gold'' label over a ``silver'' label.  Notably, we do not take into account the goodness of fit of the magnetar model itself when determining labels.\footnote{We exclude points ${>}500$ rest-frame days after explosion, since MOSFiT does not allow for a variable $\gamma$-ray opacity, which changes the light-curve slope at very late times. This only affects SN~2015bn.} Nonetheless, we can conclude that a significant fraction of hydrogen-poor SLSNe deviate from the magnetar model after peak.

\subsection{Bump Masking\label{sec:model2}}
\begin{figure*}
    \begin{minipage}{0.5\textwidth}
        \includegraphics[width=\textwidth]{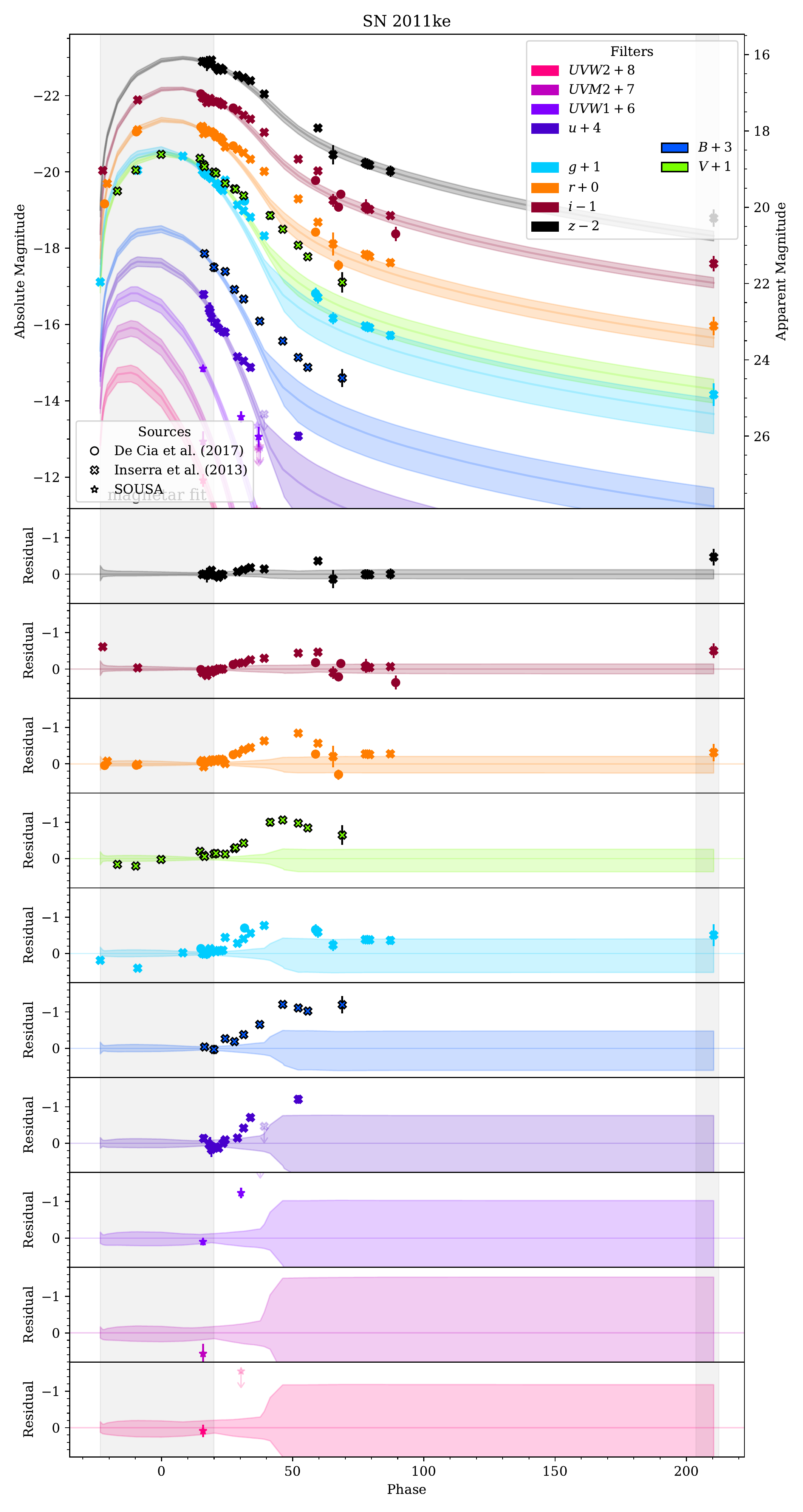}
    \end{minipage}
    \begin{minipage}{0.5\textwidth}
        \includegraphics[width=\textwidth]{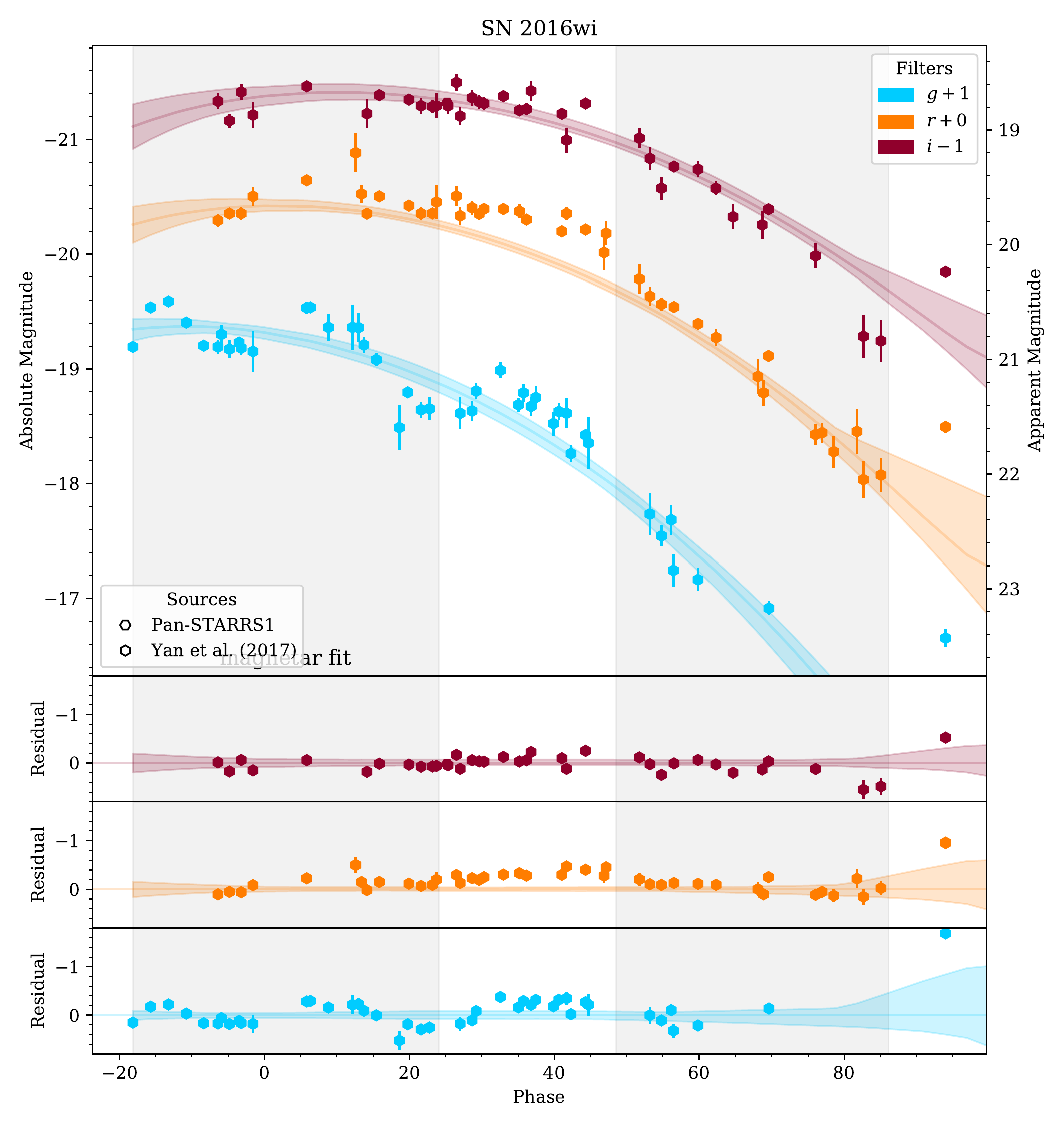}
        \includegraphics[width=\textwidth]{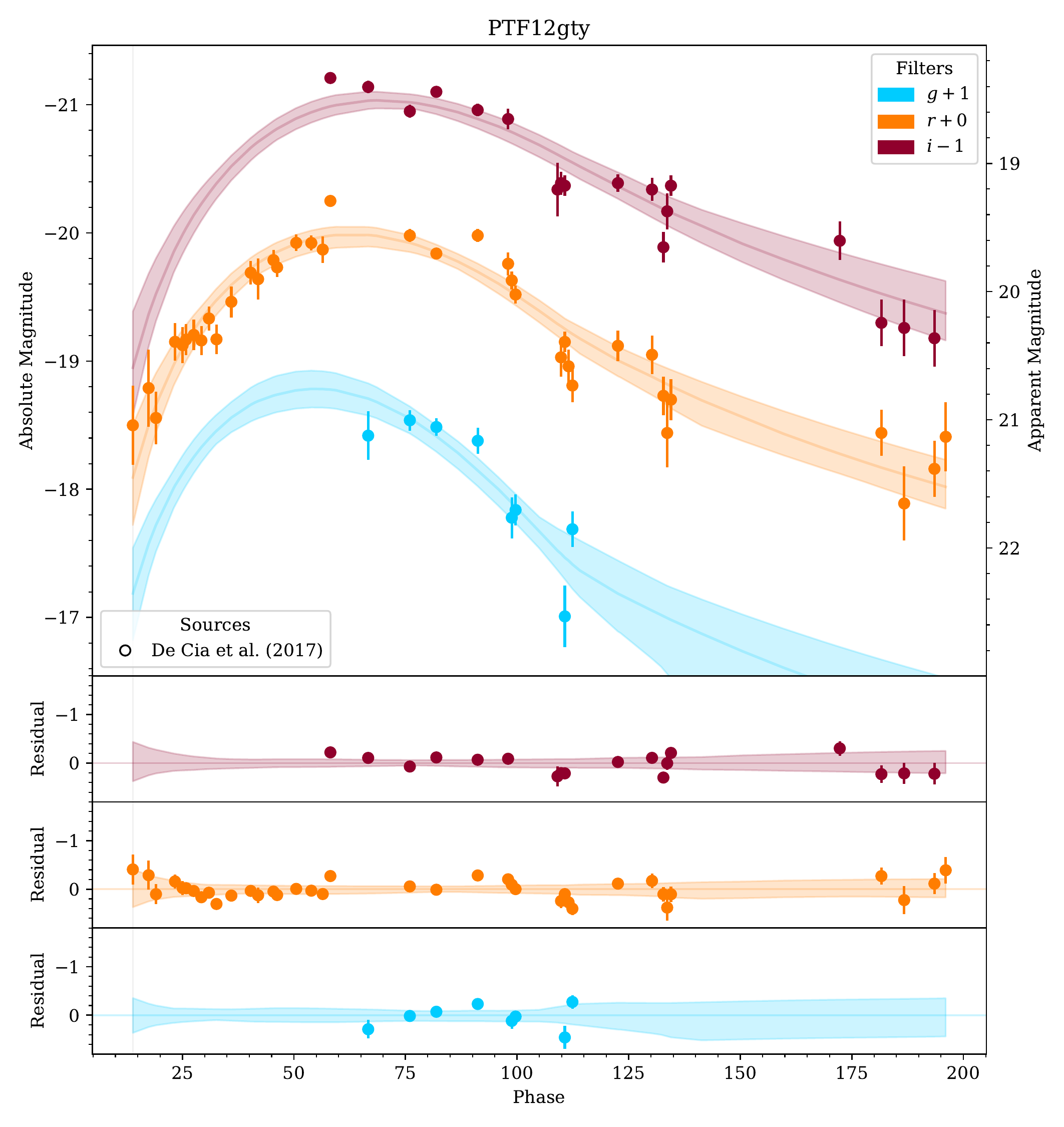}
    \end{minipage}
    \caption{Left: The light curve of SN~2011ke (gold sample) compared to its MOSFiT model, with residuals. Points outside the gray shaded region are converted to upper limits during the fit in an attempt to model only the contribution of the underlying engine. With respect to this model, the light curve shows a significant excess in all observed filters, peaking ${\sim}50$ rest-frame days after maximum light. Top right: Same plot for SN~2016wi (silver sample), showing a possible excess ${\sim}40$ rest-frame days after maximum light. Bottom right: Same plot but for PTF12gty (control sample), showing no significant excess at any observed phase.\\(The complete figure set (34 images) is available.)}
    \label{fig:mosfit}
\end{figure*}

For events in the gold and silver samples, observations during the bump may be affecting the fit and hence the resulting model parameters. Therefore, we refit the magnetar model using only the unaffected points (e.g., the shaded region in Figure~\ref{fig:mosfit}), treating all remaining points as upper limits, and use these models for the remainder of our analysis. This allows us to exclude models that are brighter than the observed bump, under the assumption that the bump is a purely additive power source on top of the underlying magnetar model. In principle, the light-curve undulations could also be ``dips'' (negative bumps), but our modeling procedure does not allow these to be identified.

In addition to the multiband model light curves, we calculate a pseudobolometric model light curve by implementing a filter in MOSFiT whose transmission function is flat between 347 and 887\,nm (the blue edge of the $U$ filter to the red edge of the $I$ filter). We restrict ourselves to this range because most SLSNe in the literature do not have ultraviolet or infrared light curves, which means the true bolometric luminosity is not well constrained over their entire evolution. Likewise, we calculate pseudobolometric observed light curves by fitting the modified-blackbody spectral energy distribution (SED) of \cite{nicholl_magnetar_2017} to each epoch of multiband photometry\footnote{For CSS130912 and iPTF13cjq, which show a bump when only one photometric band is observed, we include epochs with only a single band by using the temperature posterior from the previous epoch as the prior for the new epoch. The $T_\mathrm{bump}$ measurements will therefore not be reliable for these two events.} and integrating between the $U$ and $I$ bands. These observed and model pseudobolometric light curves should be directly comparable (Figure~\ref{fig:mosfit_bol}, top). We then subtract the pseudobolometric model light curve (from MOSFiT) from the pseudobolometric observations (from SED integration) and normalize to the model to examine the fractional residuals during the bumps (Figure~\ref{fig:mosfit_bol}, center).

\subsection{Bump Modeling\label{sec:model3}}
\begin{figure*}
    \centering
    \includegraphics[width=\columnwidth]{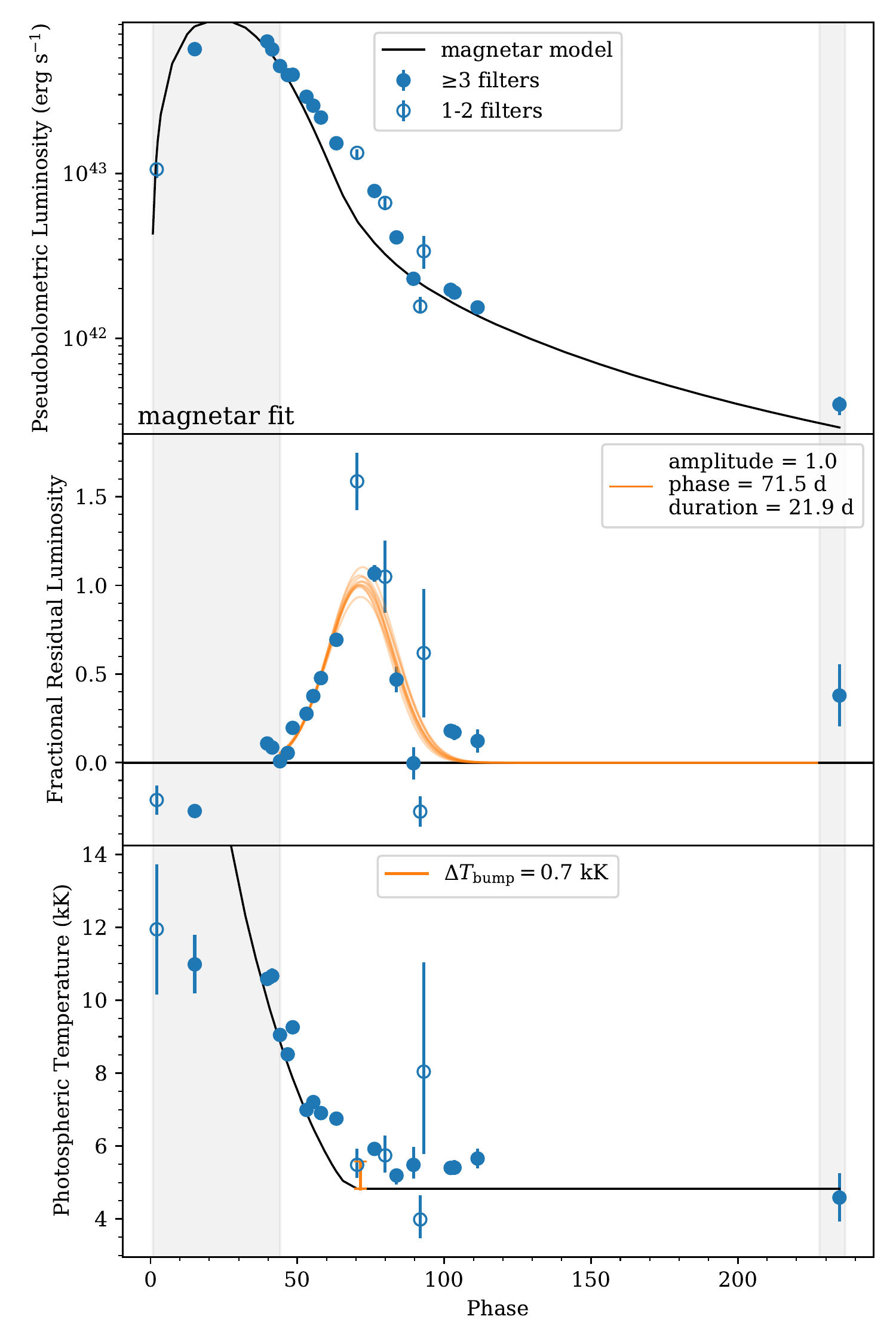}
    \includegraphics[width=\columnwidth]{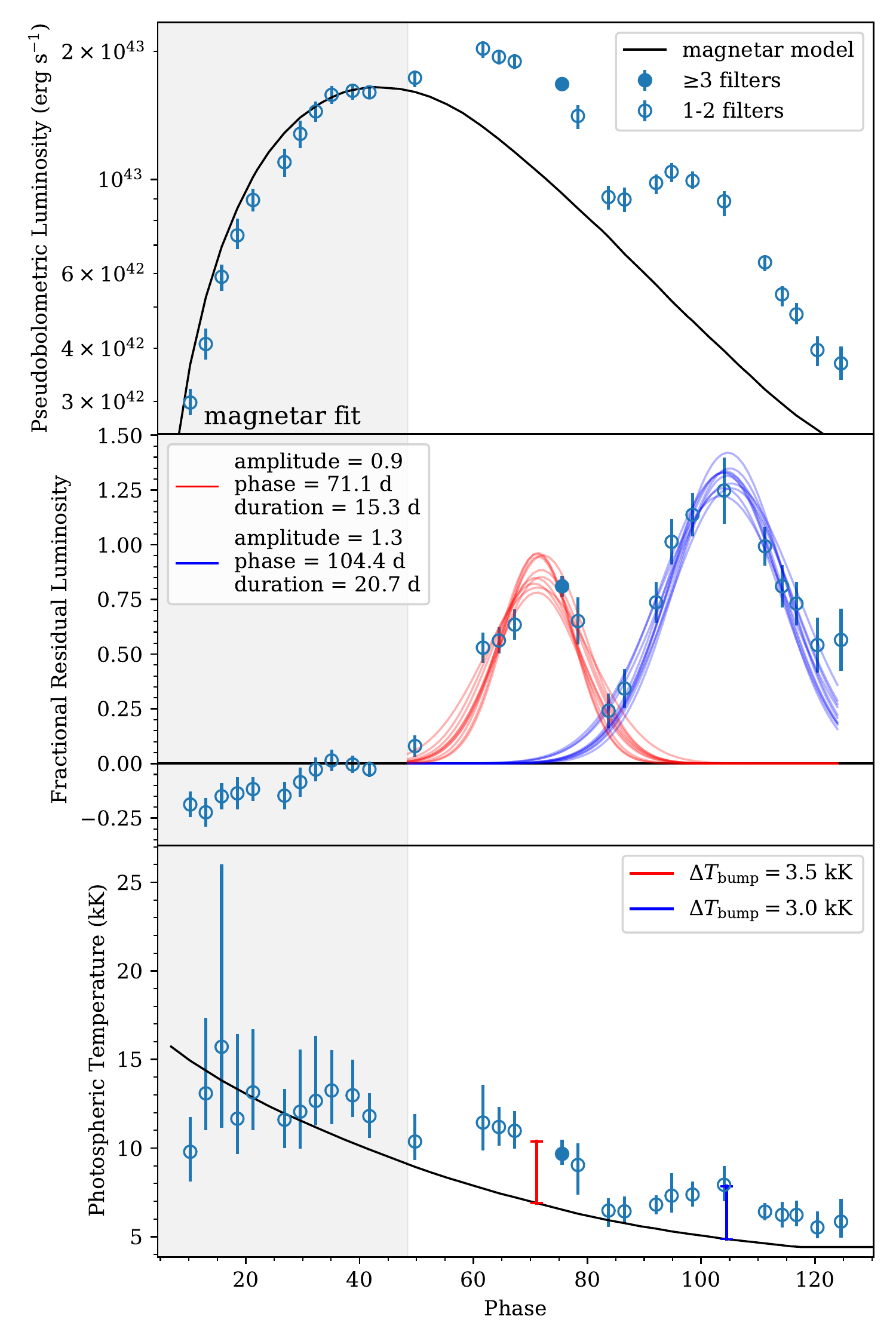}    \caption{Top left: The pseudobolometric ($U$ to $I$) light curve of SN~2011ke compared to its magnetar model. Open points are calculated from SEDs with only two observed filters. Center left: Fractional residuals fit with a Gaussian (orange lines) to determine the approximate amplitude, phase, and duration of the excess. The gray shaded region shows the time range used to determine the magnetar model. Right: Same plot but for SN~2019hge, which shows two distinct bumps.\\(The complete figure set (26 images) is available.)}
    \label{fig:mosfit_bol}
\end{figure*}

Lastly, we fit a Gaussian to these fractional residuals in order to characterize the amplitude, phase, and FWHM duration of the excess (Figure~\ref{fig:mosfit_bol}, center). In addition to the amplitude and phase of the bump, we calculate the blackbody temperature at the peak of the excess by linearly interpolating between SED fits to the two nearest epochs observed photometry (Figure~\ref{fig:mosfit_bol}, bottom). We assume a 10\% uncertainty on the temperature for the purposes of calculating the correlation coefficients in Section~\ref{sec:correlations}.

In a handful of cases (marked with footnotes in Table~\ref{tab:sne}), the residuals seem to indicate two distinct bumps. For these SNe, we follow the same procedure but simultaneously fit two Gaussians to the residuals (Figure~\ref{fig:mosfit_bol}, right). These bumps are treated as independent data points in the rest of our analysis.

\section{Analysis\label{sec:analysis}}
\subsection{Bump Prevalence\label{sec:prevalence}}
\begin{figure*}
    \centering
    \includegraphics[width=\textwidth]{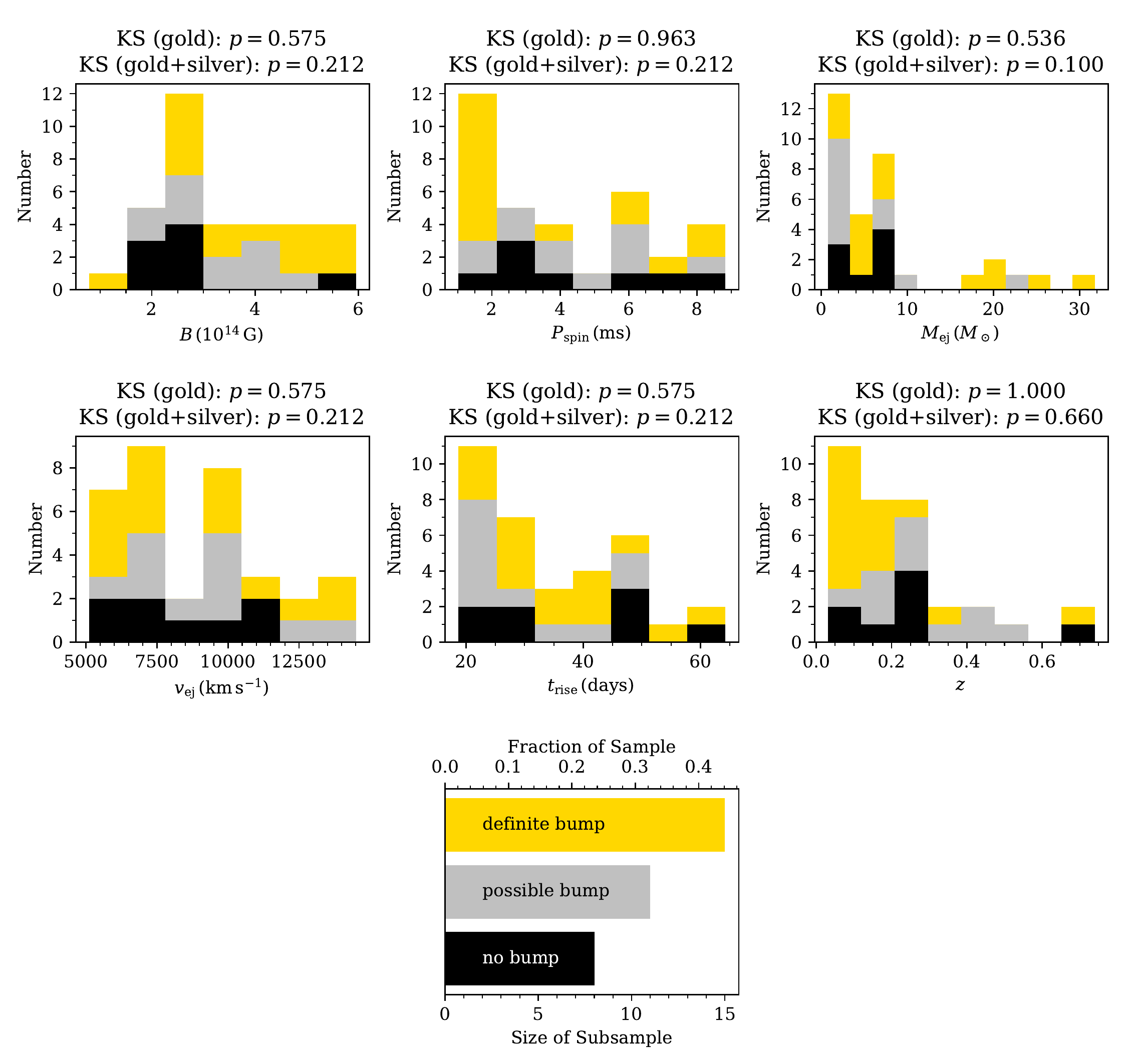}
    \caption{Stacked histograms comparing the SN properties for the gold, silver, and control samples of SLSNe. For each parameter, we perform Kolmogorov--Smirnov tests comparing the gold sample to the control and the combined gold+silver sample to the control. The $p$-values are reported above each panel. Unfortunately, the small number of the control sample does not allow us to discern any statistically significant differences. The bottom right panel compares the sizes of the three subsamples.\\(The data used to create this figure are available.)}
    \label{fig:hists}
\end{figure*}

Our main conclusion from this work is that the majority of SLSNe have noticeable excesses relative to the magnetar model in the months after peak. Figure~\ref{fig:hists} (bottom right) illustrates the relative sizes of the gold, silver, and control samples we define above. Depending on whether we include the silver sample, between \goldpct{}\% and \bumpspct{}\% of SLSNe show an excess. As we caution above, the assignment of any of these qualitative labels to a given SN can be disputed, but the broader conclusion is robust.

We are therefore searching for a physical process that is common among SLSN progenitors that can radiate significant energy in addition to the magnetar model, rather than a rare process that produces peculiar SLSNe. This process may in fact be ubiquitous among SLSNe if it were possible to observe them all out to late phases. Notably, three of the best observed SLSNe in the literature, SNe~2015bn, 2017egm, and 2017gci, all show multiple bumps in their light curves thanks to their extensive phase coverage.

However, it is worth examining what observational biases could affect our results. For example, the detection of a bump suffers from a version of the \cite{malmquist_relations_1922,malmquist_contribution_1925} bias: SNe with a bump are, by definition, brighter for a longer period of time, and thus are more likely to be detectable at the phases where the bump occurs. SNe without a bump will fade more quickly and therefore not be eligible for inclusion in our control sample. Similarly, we are more likely to detect bumps at earlier phases than at later phases, all else being equal. Because the SNe in our sample were discovered and followed up by a wide range of telescopes, it is not straightforward to quantitatively assess the effect of this bias on our results.

\subsection{Sample Comparisons\label{sec:compare}}
For each SN in our sample, we consider five ``SN properties'' as determined by MOSFiT modeling---the magnetar magnetic field ($B$) and spin period ($P_\mathrm{spin}$), the ejecta mass ($M_\mathrm{ej}$) and velocity ($v_\mathrm{ej}$), and the rise time of the rest-frame pseudobolometric model light curve ($t_\mathrm{rise}$)---as well as the SN redshift ($z$). Figure~\ref{fig:hists} plots the distributions of these parameters for each of our three subsamples.

We investigate whether there is a difference between the properties of SNe with and without a bump by performing two-sample \cite{kolmogorov_sulla_1933}--\cite{smirnov_table_1948} tests comparing the gold sample to the control sample and the combined gold+silver sample to the control sample. The resulting $p$-values are shown above each panel in Figure~\ref{fig:hists}. We do not see any obvious differences between our three subsamples. However, the statistical significance of these tests is limited by the smaller sample size, in this case the eight SNe in the control sample.

\begin{figure*}
    \centering
    \includegraphics[width=\textwidth]{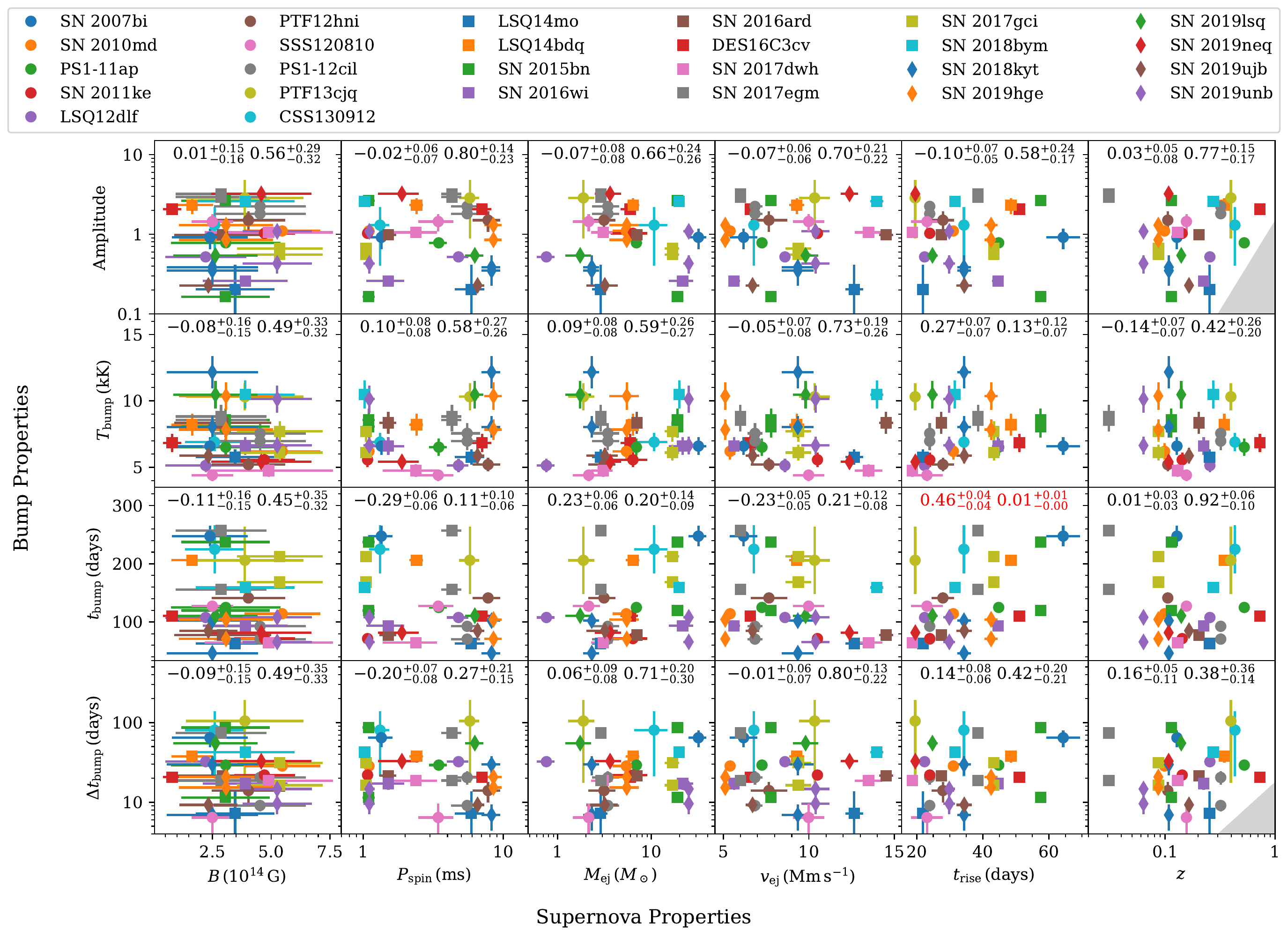}
    \caption{Correlations between the SN/progenitor properties (as determined by MOSFiT; horizontal axes) and the properties of the bump (vertical axes) for each event in our sample. The numbers at the top of each panel are the Spearman rank correlation coefficient and its $p$-value, with $1\sigma$ uncertainties. The strongest significant correlation is highlighted with red text and discussed in Section~\ref{sec:correlations}. Gray regions mark parts of parameter space unlikely to be observed due to Malmquist bias.\\(The data used to create this figure are available.)}
    \label{fig:correlations}
\end{figure*}

\subsection{Correlations\label{sec:correlations}}
In addition to the aforementioned SN properties, for each SN in the gold and silver samples, we consider four ``bump properties'': the amplitude (as a fraction/multiple of the magnetar-powered luminosity); the modified-blackbody temperature at the peak of the bump, $T_\mathrm{bump}$; the phase of the bump, $t_\mathrm{bump}$ (in rest-frame days after explosion); and the duration of the bump, $\Delta t_\mathrm{bump}$ (FWHM in rest-frame days). Figure~\ref{fig:correlations} plots the bump properties against the SN properties for each event.

We search for correlations between the SN properties and the bump properties using the \cite{spearman_proof_1904} rank correlation coefficient, which is sensitive to any monotonic relationship, linear or not. In order to take uncertainties on the parameters into account, we use a Monte Carlo routine to perturb each point within its posterior (approximated as a Gaussian) and remeasure the correlation coefficient 100 times, following the method of \cite{curran_monte_2014}. Each panel of Figure~\ref{fig:correlations} shows the correlation coefficient ($\rho$) followed by the corresponding $p$-value, each with $1\sigma$ uncertainties from the Monte Carlo routine. Large $|\rho|$ and small $p$ indicate strong and statistically significant correlations, respectively.

\begin{figure}
    \includegraphics[width=\columnwidth]{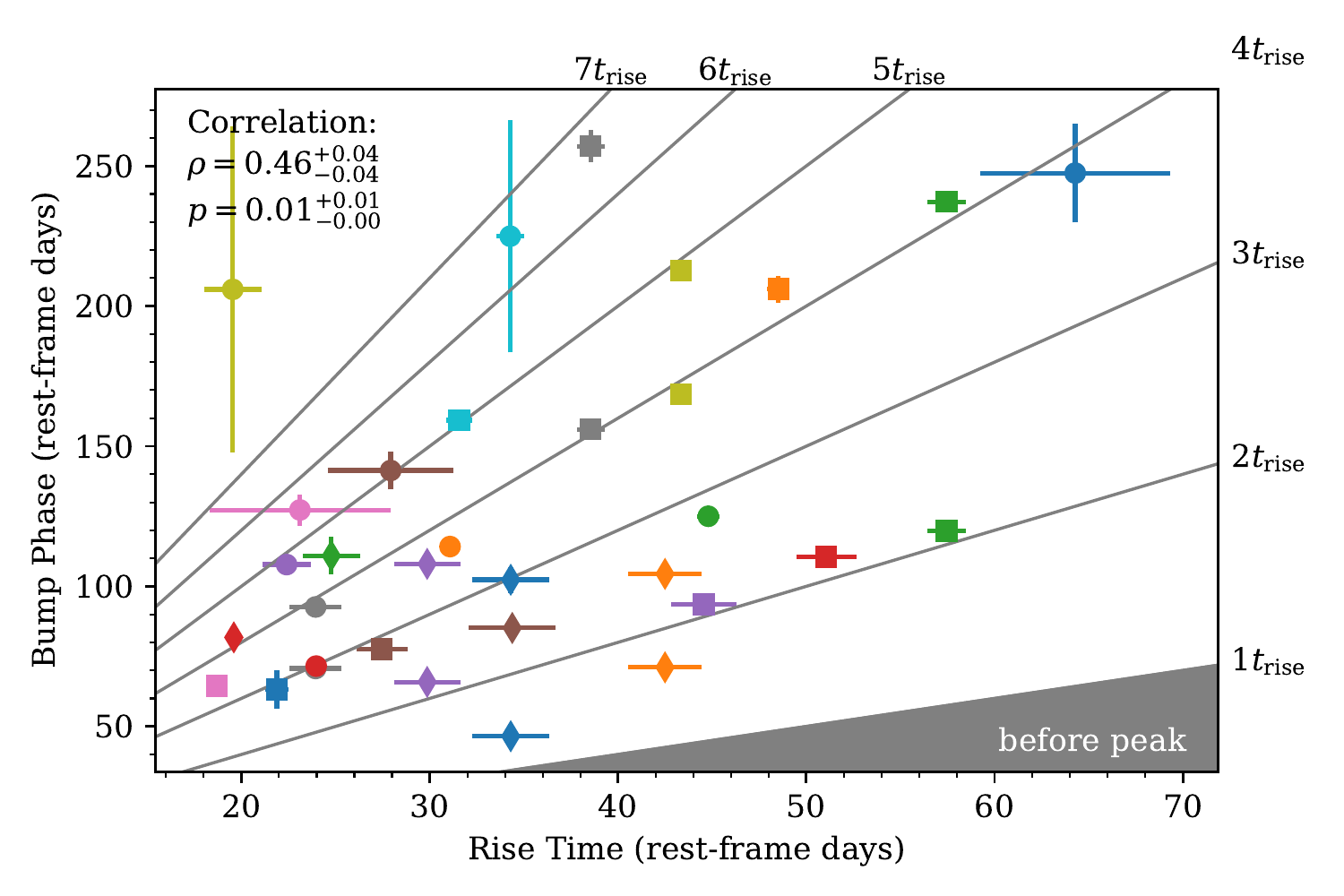}
    \includegraphics[width=\columnwidth]{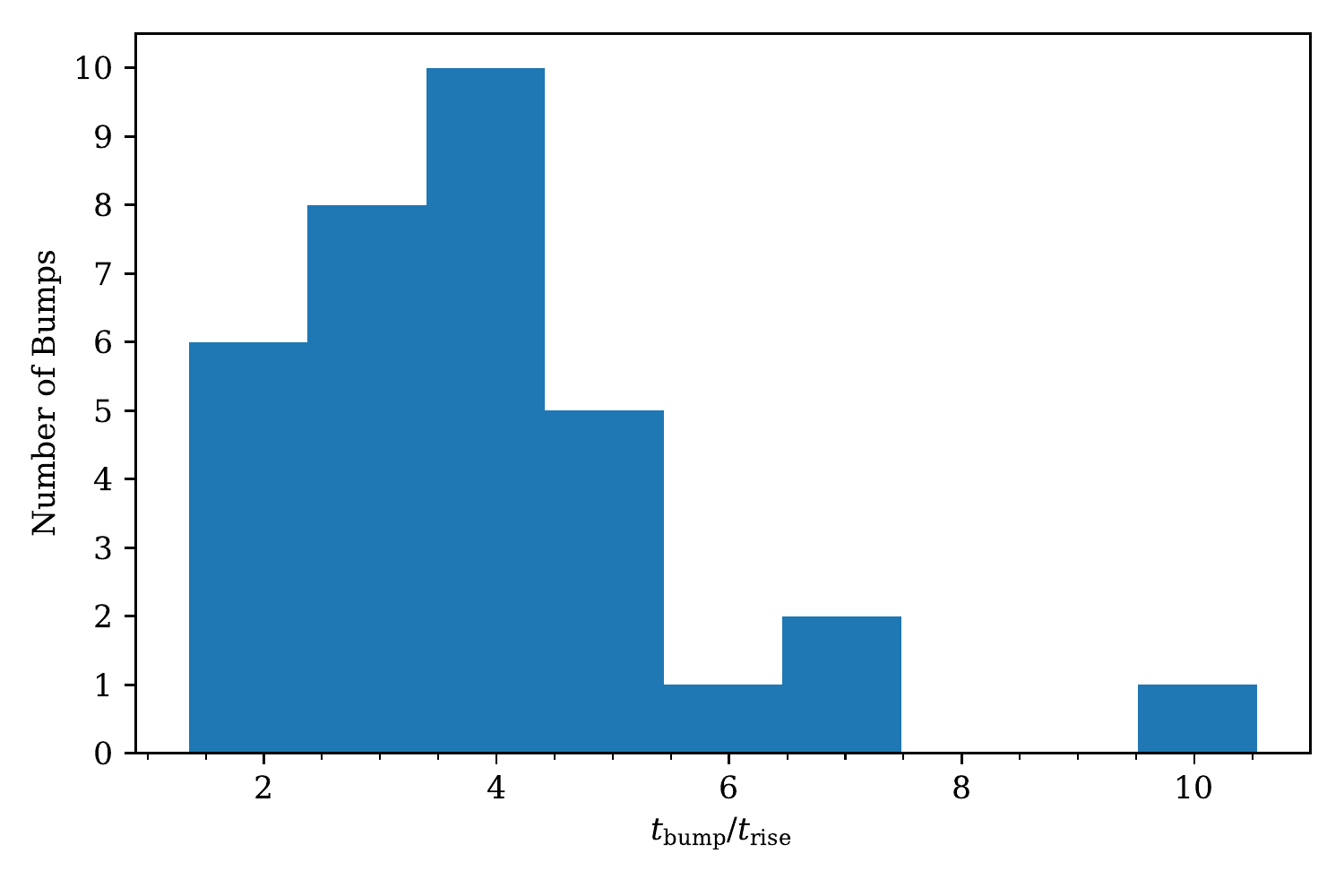}
    \caption{Top: A larger view of the strongest correlation in Figure~\ref{fig:correlations}. The correlation between the bump phase and the rise time can be understood as a scarcity of bumps happening at $t_\mathrm{bump} \lesssim 2 t_\mathrm{rise}$ (the bottom right portion of this plot). If the source of variability is centrally located, it would take approximately this long for the photons it produced to escape the ejecta. There is also an observational bias against finding bumps at very late phases in rapidly evolving SNe (the upper left portion of this plot). Bottom: The distribution of bump phases normalized by the rise time of each SLSN. The positive correlation discussed above implies that this distribution is narrow compared to the raw distribution of bump phases.\label{fig:bumpphase_risetime}}
\end{figure}

We find a moderate positive correlation ($\rho = 0.46 \pm 0.05$) between the phase of the bump (explosion to bump maximum) and the rise time (explosion to magnetar maximum), which suggests that bumps tend to occur at a specific ``evolutionary phase,'' namely $t_\mathrm{bump} = (3.7 \pm 1.4) t_\mathrm{rise}$ after explosion. Figure~\ref{fig:bumpphase_risetime} shows that this correlation can be understood as a scarcity of bumps happening at $t_\mathrm{bump} \lesssim 2 t_\mathrm{rise}$, or one rise time after peak. (Bumps happening before maximum light are already excluded by our selection criteria, although in principle this analysis could also be applied to pre-peak bumps.) This is approximately how long it would take for centrally produced photons to escape the ejecta, which we discuss further in Section~\ref{sec:depth}. In addition, we are biased against observing bumps at late phases in rapidly evolving (short rise time) SNe because those SNe will be faint at those phases, similar to the bias against discovering rapidly evolving SNe discussed by \cite{nicholl_diversity_2015}.

Somewhat surprisingly, we do not see a correlation with redshift in any of the bump properties. (Relatedly, our second-highest-redshift event is in the gold sample; see Figure~\ref{fig:hists}.) Nonetheless, we do observe a lack of weak and short bumps at high redshift (gray regions in Figure~\ref{fig:correlations}). These may indicate an observational bias that becomes important for $z \gtrsim 0.3$. Therefore, we repeat our correlation analysis for events with $z < 0.3$ only. We still find a moderate correlation between phase and rise time ($\rho = 0.50 \pm 0.04$ and $p \approx 0.01$), and no other correlations become significant (the strongest being $T_\mathrm{bump}$ vs.\ $t_\mathrm{rise}$ with $\rho = 0.41 \pm 0.07$ and $p=0.04_{-0.02}^{+0.05}$). This indicates that observational biases are not a limiting factor in our analysis. Nonetheless, these biases will exist in any uniformly observed sample and must be considered in future work.

\vspace{1in}
\subsection{Depth of Luminosity Source\label{sec:depth}}
Photon diffusion acts as a low-pass filter on any variable luminosity produced inside optically thick SN ejecta. Sudden changes in the input luminosity are both delayed and smeared out before reaching the observer. Therefore, we can use the phases and durations of the bumps in our sample to place limits on the depth at which they were produced.

The diffusion time through spherical ejecta with opacity $\kappa$ from a radius $r$ to the outer edge $R$ is given by
\begin{equation}
    t_\mathrm{diff}(r, R) = \frac{\kappa}{c} \int_r^R \rho(r) r dr.
\end{equation}
Given a standard radial density profile within the core of the ejecta \citep{chevalier_asymmetric_1989},
\begin{equation}
    \rho(r) = \frac{M}{2\pi R^2 r},
\end{equation}
where $M$ is the total ejecta mass, the diffusion time decreases linearly with radius:
\begin{equation}
    t_{\mathrm{diff}}(r, R) = \frac{M \kappa}{2\pi c R} \left(1 - \frac{r}{R} \right).
\end{equation}
For ejecta in homologous expansion, we can rewrite this in terms of the dimensionless depth below the surface of the ejecta, $\delta \equiv 1 - \frac{r}{R}$, and the elapsed time since explosion, $t = \frac{R}{v}$, where $v$ is the velocity of the outer edge of the ejecta:
\begin{equation}
    t_{\mathrm{diff}}(\delta, t) =  \frac{M \kappa \delta}{2\pi c v t}.
\end{equation}
The light curve peaks when the elapsed time matches the diffusion time from the center of the explosion ($\delta=1$), so the rise time is given by
\begin{equation}
    t_\mathrm{rise} = t_{\mathrm{diff}}(1, t_\mathrm{rise}) = \frac{M \kappa}{2\pi c v t_\mathrm{rise}}.
    \label{eq:risetime}
\end{equation}
Applying this boundary condition, we get
\begin{equation}
    t_{\mathrm{diff}}(\delta, t) = \frac{t_\mathrm{rise}^2 \delta}{t}.
\end{equation}

\begin{figure}
    \includegraphics[width=\columnwidth]{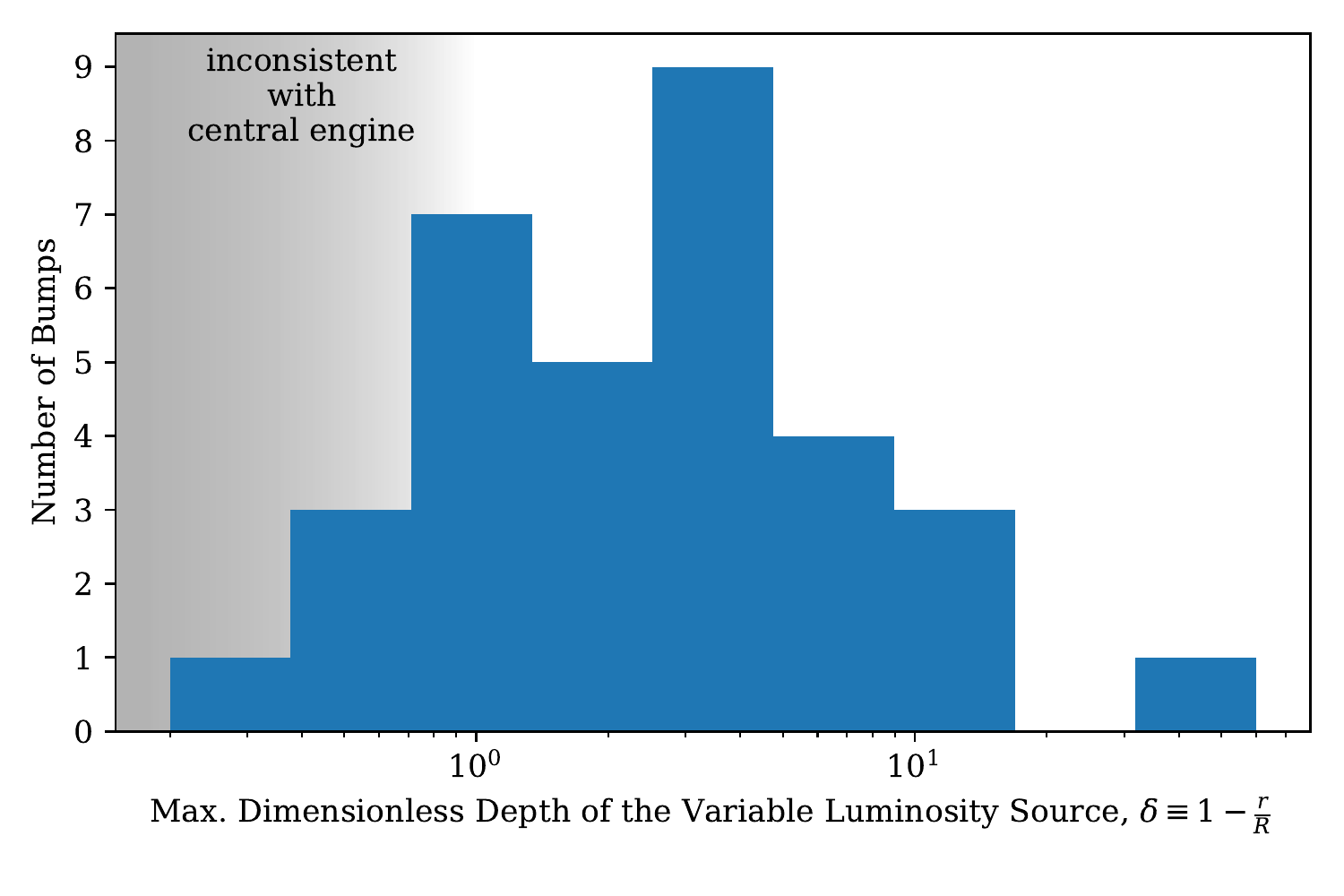}
    \includegraphics[width=\columnwidth]{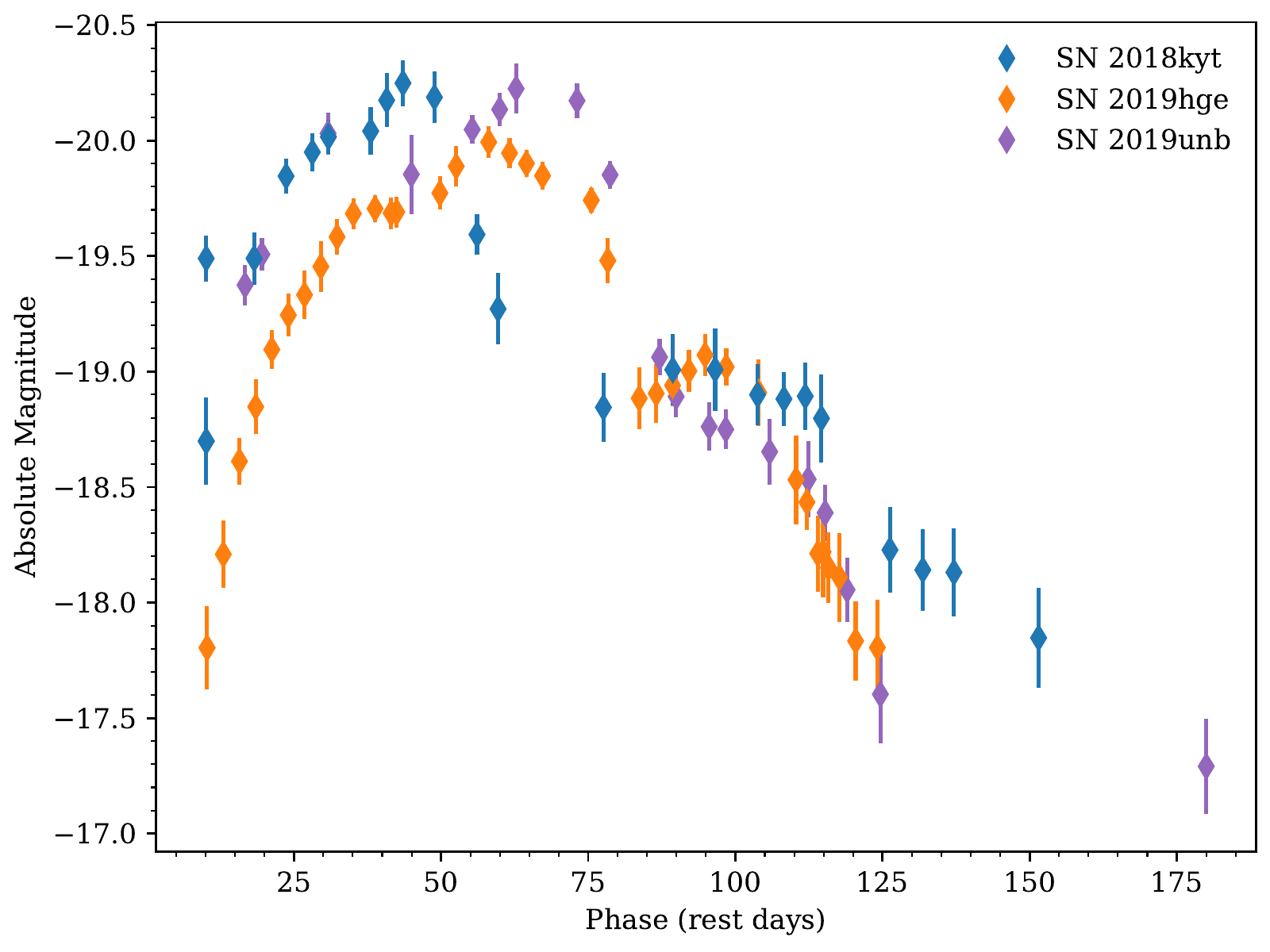}
    \caption{Top: The maximum dimensionless depth (Eq.~\ref{eq:depth}) at which the bumps in our sample could have been produced. Bumps in the shaded region could not have been produced by the central engine. However, this should only be considered as an order-of-magnitude calculation.\label{fig:depth} Bottom: SNe~2018kyt, 2019hge, and 2019unb all have two distinct bumps in their light curves (the $g$ band plotted here), with similar light-curve morphologies and similar, relatively faint, peak magnitudes. The initial bumps in these SNe are among the few in our sample that are inconsistent with a central engine origin (gray region in top panel). All three also belong to the sample of helium-rich SLSNe of \cite{yan_helium-rich_2020}. Together, these pieces of evidence may suggest a CSM origin for this type of bump.\label{fig:multibump}}
\end{figure}

If we observe a bump at time $t_\mathrm{bump}$, its duration must be at least the diffusion time through the material along the line of sight:
\begin{equation}
    \Delta t_\mathrm{bump} \gtrsim t_\mathrm{diff}(\delta, t_\mathrm{bump}) = \frac{t_\mathrm{rise}^2 \delta}{t_\mathrm{bump}}.
\end{equation}
This means we can place an upper limit on the depth at which it was produced:
\begin{equation}
    \delta \lesssim \frac{t_\mathrm{bump} \Delta t_\mathrm{bump}}{t_\mathrm{rise}^2}.
\label{eq:depth}
\end{equation}
Specifically, if the right-hand side of Eq.~\ref{eq:depth} is significantly less than unity, we can rule out that the variability was produced at the center of the ejecta. We calculate this quantity for each bump in our sample and plot the results in Figure~\ref{fig:depth}. However, this calculation is very approximate (e.g., defining $\Delta t_\mathrm{bump}$ as the FWHM rather than as any other measure of duration) and should only be treated as an order-of-magnitude result.

The majority of bumps have depths greater than or consistent with 1. This allows for all possibilities: that the variability is generated by the central engine early in the SN evolution, or that the variability is generated later on in the outer layers of the ejecta, as would be the case for circumstellar interaction. However, several of the bumps have a depth of $\delta \lesssim 1$. One is the ``knee'' of SN~2015bn, which has been discussed by \cite{nicholl_sn_2016}. Three other such light curves are shown in the bottom panel of Figure~\ref{fig:multibump}. These SNe share several characteristics: (1) they have similar light-curve morphologies, with an initial bump immediately after the underlying peak, followed by a secondary bump ${\sim}40{-}60$ days later. (2) Their underlying peaks are relatively faint for SLSNe, $M_g \gtrsim -20$~mag. (3) Unlike most SLSNe, \cite{yan_helium-rich_2020} showed that these events (and three others) show helium features in their spectra. The helium features, along with the exclusion of a central origin for the light-curve bumps, may suggest that this specific type of bump is produced by circumstellar interaction. However, it may not be possible to generalize this conclusion to other events.

\section{Interpretation\label{sec:interp}}
We consider two possible explanations for the origins of post-peak bumps in SLSN light curves. In one case, interaction between the SN ejecta and a shell of CSM converts a fraction of the ejecta kinetic energy into luminosity over a short period of time. This is the interpretation advanced by \cite{inserra_complexity_2017}, \cite{li_energy_2020}, and others. In the second case, the bump is intrinsic to the magnetar and the SN ejecta: a sudden increase in the input luminosity or a sudden decrease in opacity could allow a burst of magnetar luminosity to escape the ejecta quickly. The latter is the interpretation advocated by \cite{metzger_ionization_2014}. (\cite{nicholl_sn_2016} consider both possibilities.) In the following subsections, we consider what clues the analysis above provides in distinguishing between these two scenarios, as well as what the consequences of each scenario would be.

\begin{figure}
    \centering
    \includegraphics[width=\columnwidth]{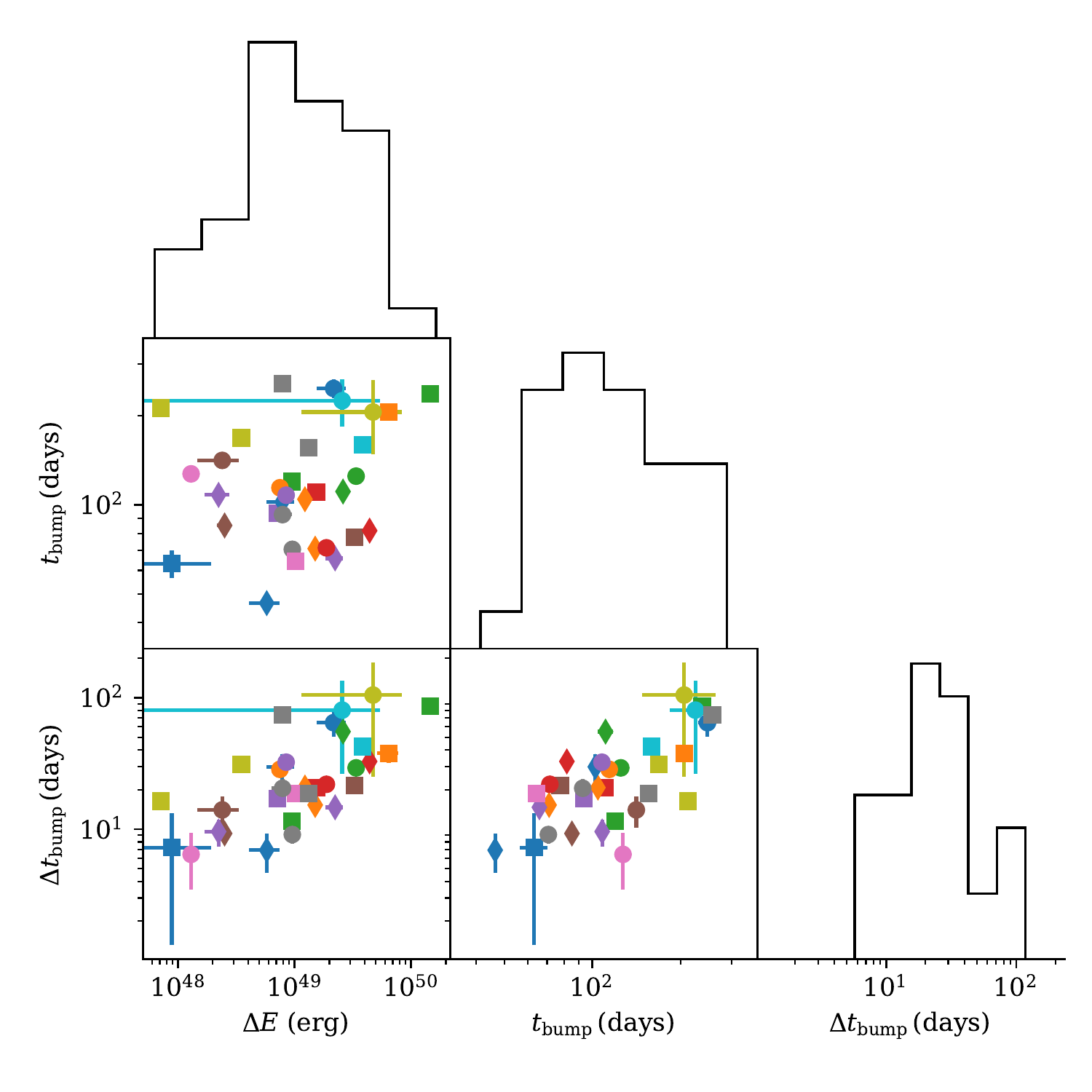}
    \includegraphics[width=\columnwidth]{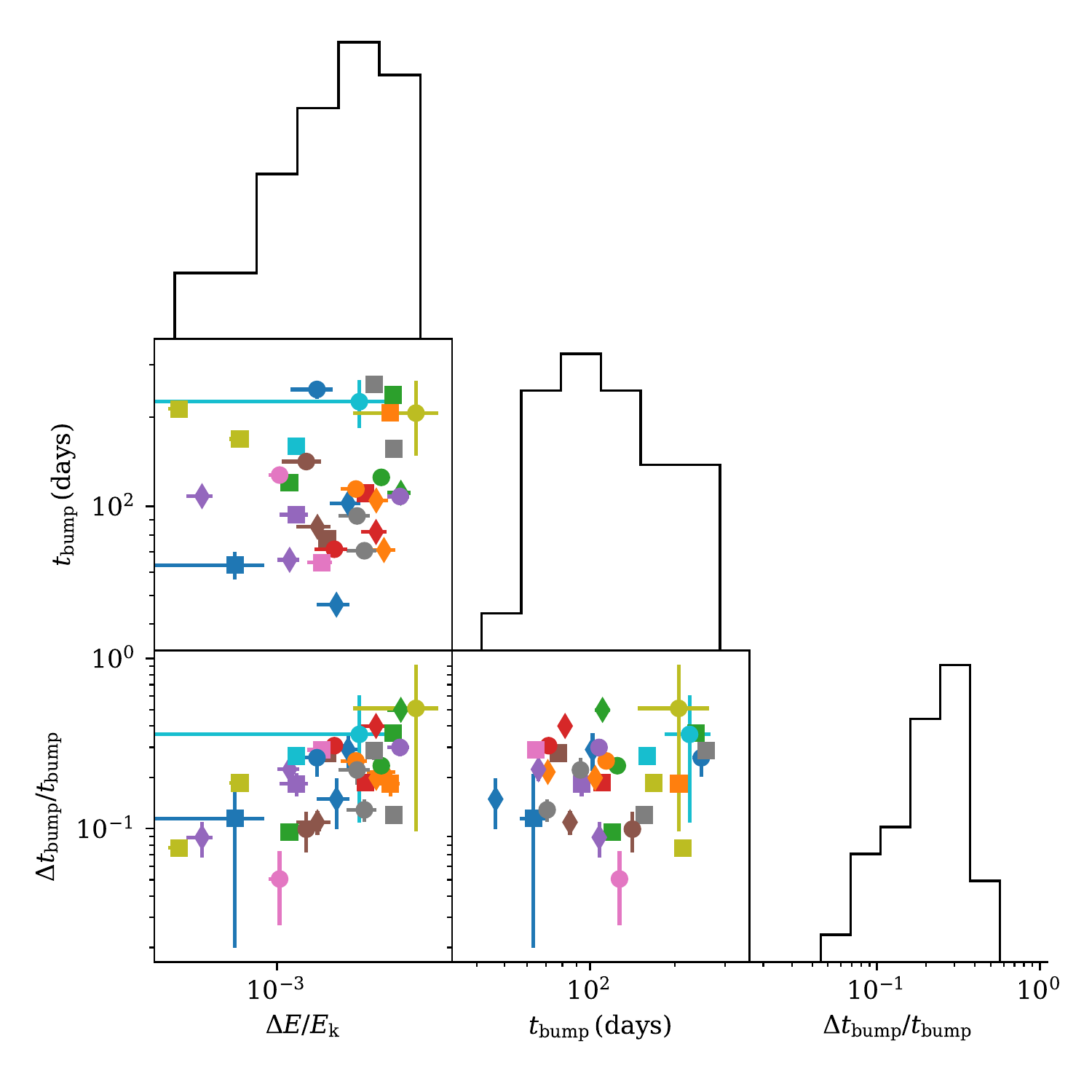}
    \caption{The phase ($t_\mathrm{bump}$), duration ($\Delta t_\mathrm{bump}$), and excess energy ($\Delta E$) of the SLSN light-curve bumps in our sample. The top panel is in physical units. In the bottom panel, the duration is given in units of the phase, and the energy is given in units of the ejecta kinetic energy.\label{fig:energy}}
\end{figure}

To estimate the amount of energy released during a bump, we integrate the Gaussian models described in Section~\ref{sec:model2} over the duration of the bump (i.e., the range of phases excluded from the second round of MOSFiT modeling). In the case of multiple bumps, we integrate each Gaussian model separately to determine the energy released in each bump. The distribution of excess radiated energies is plotted in Figure~\ref{fig:energy}. We find that these bumps emit $(0.26{-}3.38) \times 10^{49}$~erg (throughout this section we list the 16th and 84th percentiles of the sample), which corresponds to $0.14{-}1.94\%$ of the ejecta kinetic energy. This energy is emitted over $10{-}54$ days. These are the basic properties that any physical model has to explain.

\subsection{Circumstellar Interaction\label{sec:csm}}
\begin{figure}
    \centering
    \includegraphics[width=\columnwidth]{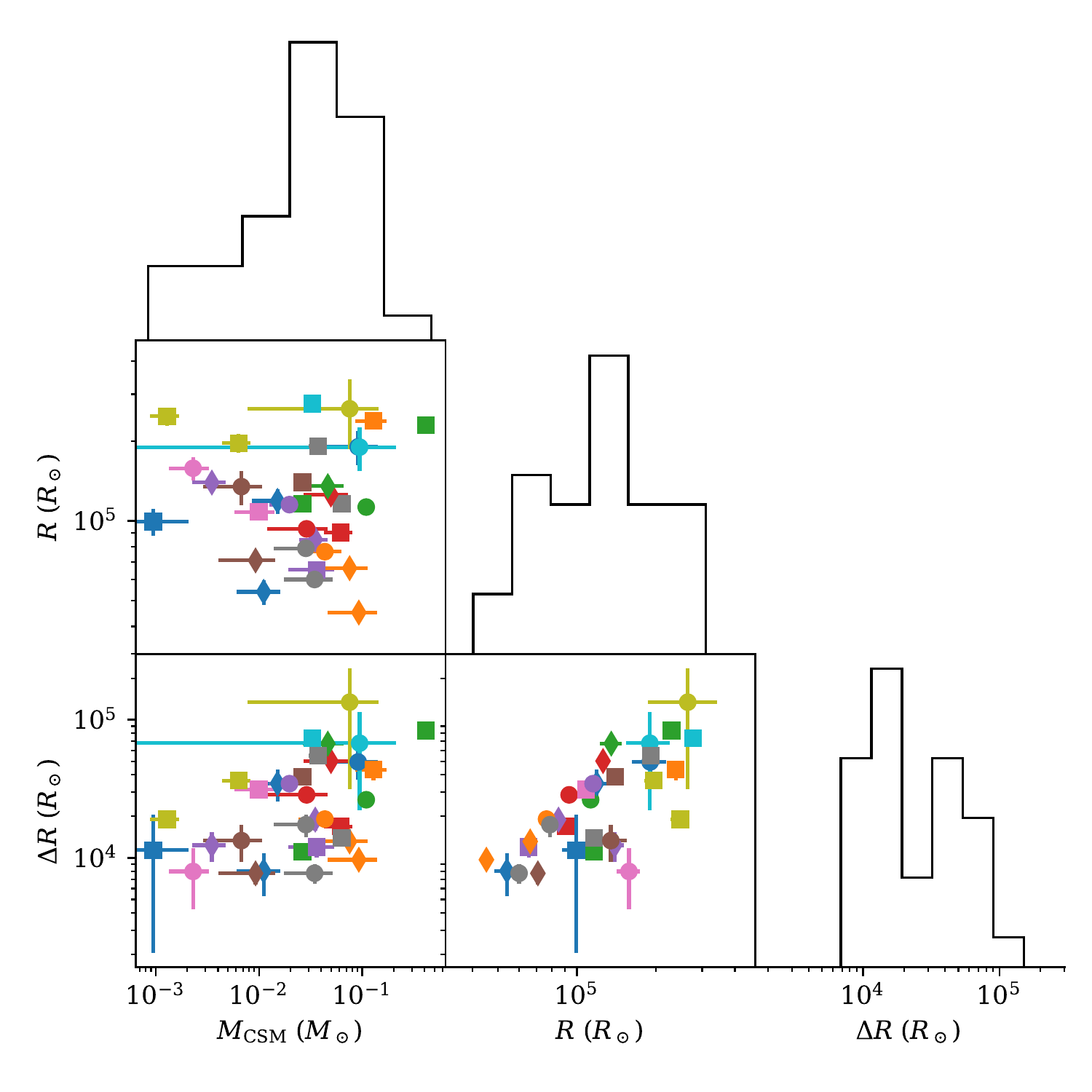}
    \includegraphics[width=\columnwidth]{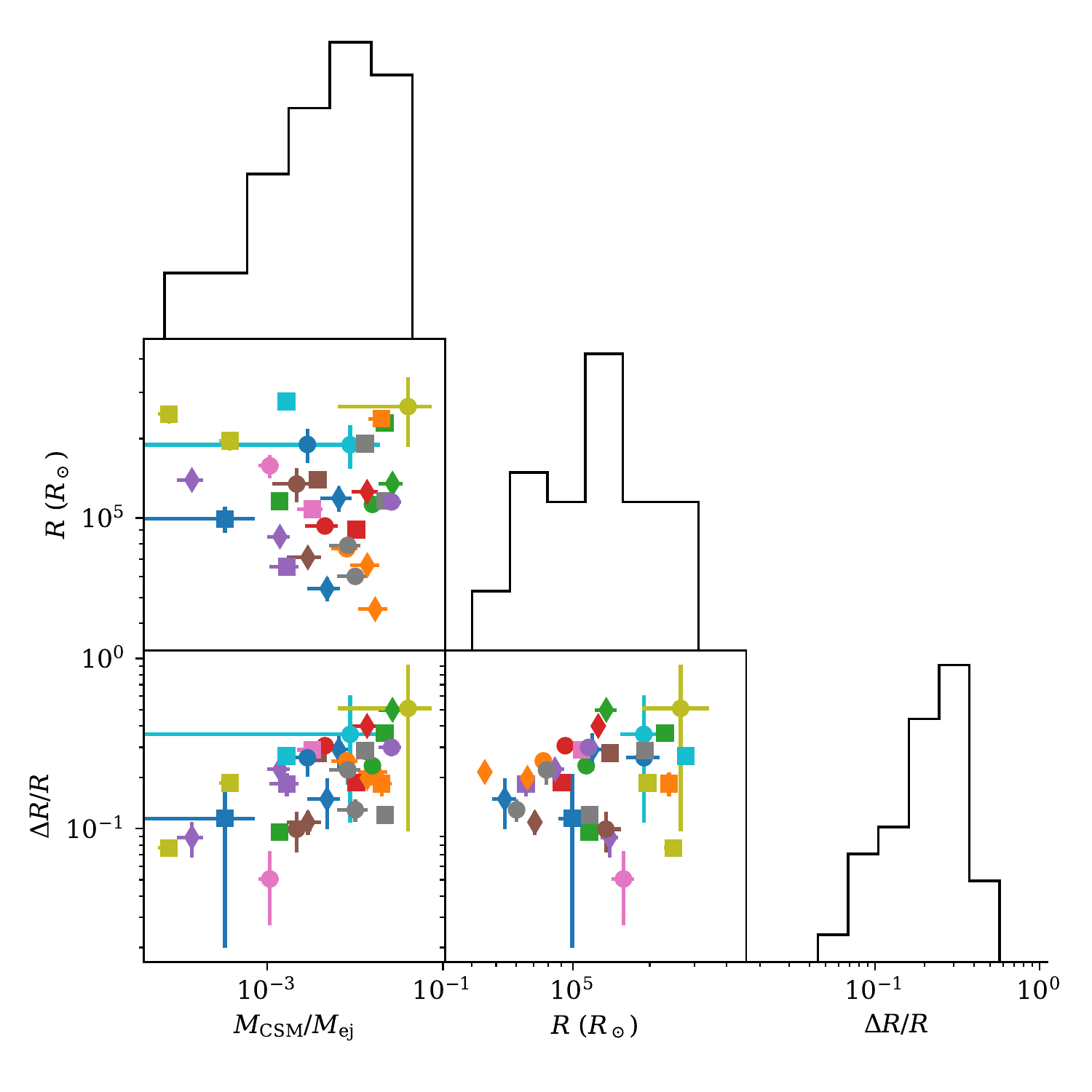}
    \caption{The radii ($R$), thicknesses ($\Delta R$), and masses ($M_\mathrm{CSM}$) of CSM shells required to explain the SLSN light-curve bumps in our sample. The top panel is in physical units. In the bottom panel, the thickness is given in units of the radius, and the mass is given in units of the ejecta mass.\label{fig:csm}}
\end{figure}

Based on the bump properties and assuming spherical symmetry, we can estimate the amount and configuration of mass required to match our observations. To estimate the mass ($M_\mathrm{CSM}$), we assume that
\begin{equation}
\frac{\Delta E}{E_\mathrm{k}} \approx \epsilon \frac{M_\mathrm{CSM}}{M_\mathrm{CSM} + M_\mathrm{ej}}
\label{eq:energy}
\end{equation}
where the left side is the fraction of the ejecta kinetic energy ($E_\mathrm{k} = \frac{3}{10} M_\mathrm{ej} v_\mathrm{ej}^2$) converted to radiation, and the right side is the fraction of kinetic energy lost in an inelastic collision, times an efficiency factor for conversion of heat into radiation that we assume to be $\epsilon = 0.5$. Under this assumption, we require $0.007{-}0.090\ M_\odot$ of CSM, which corresponds to $0.14{-}1.98\%$ of the ejecta mass (Figure~\ref{fig:csm}).

We can also estimate the radius of the CSM by calculating the radius the ejecta have reached when the bump peaks:
\begin{equation}
    R \approx v_\mathrm{ej} t_\mathrm{bump}.
\end{equation}
This yields CSM radii of $(0.71{-}1.96) \times 10^{5}\ R_\odot$, or $(5.0{-}13.6) \times 10^{15}$~cm.

We can then estimate the thickness of the shell ($\Delta R$) using the duration of the bump ($\Delta t$). There are two possibilities: If the CSM is optically thick, then the bump duration is related to the diffusion time through the CSM. However, if the CSM is optically thin, then the duration of the bump is related to the time it takes the ejecta to sweep up the entire CSM shell. We first calculate the optical depth using our previously estimated mass and radius:
\begin{equation}
\tau = \kappa \rho_\mathrm{CSM} \Delta R = \frac{\kappa M_\mathrm{CSM}}{4 \pi R^2}.
\end{equation}
We find all CSM shells to be optically thin ($\tau < \frac{2}{3}$). We can therefore approximate the shell thicknesses using
\begin{equation}
\Delta R \approx v_\mathrm{ej} \Delta t,
\end{equation}
which yields thicknesses of $(1.11{-}5.47) \times 10^{4}\ R_\odot$ (Figure~\ref{fig:csm}), or $10.9{-}30.6\%$ of the CSM radius.

The picture that emerges is of one or more thin, low-mass, low-optical-depth shells of material located ${\sim}10^5\ R_\odot$ from the progenitor star. Assuming a wind/ejection velocity of order 1000~km~s$^{-1}$, this CSM could have been produced only a few years before explosion. This also accounts for only a small fraction of the mass of the very massive stars expected to give rise to SLSNe. Overall, this scenario is plausible, given the uncertainties in massive star evolution, but we so far lack the smoking-gun evidence of narrow emission lines in SLSN spectra during their light curve bumps (see Section~\ref{sec:spec})

\cite{chatzopoulos_generalized_2012,chatzopoulos_analytical_2013} present a model of interaction-powered SNe (including SLSNe), but this model assumes that the CSM extends from the progenitor radius to infinity (i.e., the inner CSM radius is negligible and the luminosity source is centrally located). This is incompatible with our observations of a bump that only contributes luminosity at late phases. Extending this model to allow a detached shell of CSM could yield more accurate measurements of the CSM mass and distribution but is outside the scope of this work.

Above we have associated centrally located variability with central engine variability. However, if CSM is not distributed spherically, but rather in an equatorial disk, it can be encompassed by the ejecta by these late phases and can act as a central luminosity source. This ``buried'' circumstellar interaction scenario has been advanced by \cite{metzger_shock-powered_2017}, for example, in the case of luminous red novae. This would also be a mechanism to hide the spectroscopic signatures of interaction we discuss in Section~\ref{sec:spec}. The only two SLSNe with spectropolarimetry (SNe~2015bn and 2017egm) show that asymmetry of the photosphere increases with time \citep{inserra_spectropolarimetry_2016,leloudas_time-resolved_2017,bose_gaia17biu/sn_2018,saito_late-phase_2020}, which could potentially be a signature of disklike CSM in the ejecta interior. In contrast, broadband polarimetry of several other SLSNe is consistent with spherical symmetry \citep{leloudas_polarimetry_2015,cikota_testing_2018,lee_imaging_2019}.

\subsection{Central Engine Origin}
In the absence of CSM, the light-curve bumps would have to be related to the central magnetar engine. This may imply that the smooth magnetar spin-down model is an oversimplification in the presence of SN ejecta. For example, fallback accretion onto the magnetar could provide additional bursts of luminosity \citep{metzger_effects_2018}. Alternatively, a sudden change in the ejecta opacity at optical wavelengths could cause a discrete bump in the light curve without the need for an additional power source. The input magnetar luminosity could continue to be smooth, while the output luminosity reprocessed by the ejecta could be bumpy. In this context, the correlation between the phase of the bump and the rise time, where we treat the latter more generically as the evolutionary timescale of the ejecta, could indicate that bumps occur when the ejecta reach a certain temperature threshold. We note that most SLSNe show a modified-blackbody temperature of ${\sim}$6000--8000~K during the bump. This range includes the recombination temperature of singly ionized oxygen \citep{jerkstrand_nebular_2014}, which dominates SLSN ejecta by mass and therefore could produce a sudden change in opacity during recombination. Relatedly, 6000~K is approximately the temperature at which SLSNe settle in the nebular phase \citep{nicholl_nebular-phase_2019}.

The strongest evidence for an origin related to the engine is the fact that most of the bumps are consistent with the diffusion of photons from a central luminosity source (Section~\ref{sec:depth}). Although our results still allow for production further out in the ejecta at later times, the fact that the rule is mostly followed may suggest that it is a requirement. Furthermore, similar bumps have not been observed in normal-luminosity Type~Ic SNe, hinting that they can only be produced in the presence of a central engine. However, the fact that the rule is not always followed suggests that not all bumps are produced by the same mechanism.

\cite{margutti_x-rays_2017} discussed the possibility of a central engine driving ionization fronts through the ejecta in the context of ASASSN-15lh, a superluminous transient that has been variously claimed to be an SLSN \citep{dong_asassn-15lh:_2016} or a tidal disruption event \citep{leloudas_superluminous_2016}. This would increase the electron scattering opacity at optical wavelengths and decrease line blanketing at UV wavelengths, causing a shift to bluer colors and an overall increase in luminosity for a short time after the ionization front breaks out of the photosphere. To investigate this mechanism, we compare the observed blackbody temperature at the peak of the bump to the blackbody temperature from the underlying magnetar model at the same time (Figure~\ref{fig:mosfit_bol}, bottom). Most, but not all, bumps correspond to a photospheric temperature higher than the best-fit underlying model: $T_\mathrm{bump} - T_\mathrm{model}(t_\mathrm{bump}) = 1000^{+2500}_{-2300}$~K. However, this would also likely be the case if the bumps were caused by circumstellar interaction.

The phases at which we observe most of the bumps also coincide with the phase where SLSN ejecta become transparent to $\gamma$-rays \citep{vurm_gamma-ray_2021}. Even if the input luminosity were smooth, this could lead to modulations in the optical light curve by changing the fraction of luminosity that is thermalized by the ejecta. Notably, this is a subtractive effect, rather than an additive effect, meaning that our modeling procedure will not produce a valid measurement of the excess energy in the bumps.

\begin{figure*}
    \centering
    \includegraphics[width=0.88\textwidth]{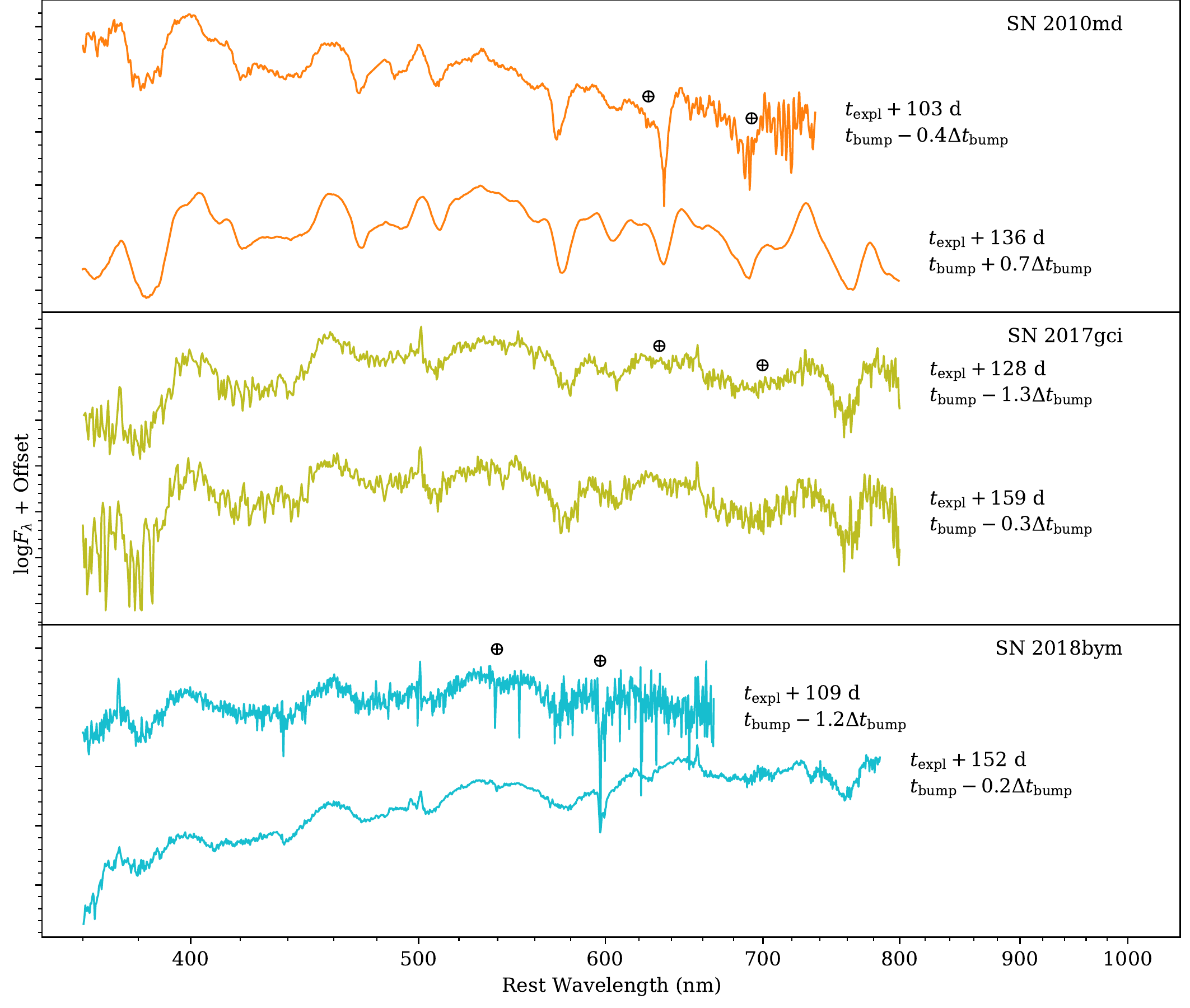}
    \caption{Three examples of SLSNe from our gold sample with spectroscopic observations both before and during their light curve bump, with their phases given in rest-frame days after explosionand in terms of the bump phase and FWHM. Telluric absorption wavelengths are marked with the $\oplus$ symbol for each SN. The details of these spectra are listed in Table~\ref{tab:bumpspec}. The spectrum of SN~2018bym at 109~d has been binned by a factor of 2 to increase signal-to-noise and to better match the resolution of the spectrum at 152~d. We do not observe a drastic change (e.g., appearance of emission lines or changes in line strengths) between any of these pairs that might explain the cause of the light curve excess.\label{fig:bumpspec}}
\end{figure*}

\begin{deluxetable*}{ccccccCc}
\tablecaption{Spectra Before and After a Light-curve Bump\label{tab:bumpspec}}
\tablecolumns{6}
\tablehead{\colhead{SN} & \colhead{MJD} & \colhead{Telescope} & \colhead{Instrument} & \colhead{Pixel Scale (nm)} & \colhead{Phase (days)} & \colhead{Phase w.r.t.\ Bump} & \colhead{Reference}}
\startdata
SN 2010md & 55436.\phn\phn\phn & WHT~4.2\,m & ISIS & 0.32 & 103.3 & -0.4 & \cite{quimby_spectra_2018} \\
SN 2010md & 55472.\phn\phn\phn & Keck~I & LRIS & 0.29--0.94 & 136.1 & +0.7 & \cite{quimby_spectra_2018} \\
SN 2017gci & 58069.213 & ESO-NTT & EFOSC & 0.51 & 128.3 & -1.3 & \cite{fiore_sn_2021} \\
SN 2017gci & 58102.201 & ESO-NTT & EFOSC & 0.51 & 158.6 & -0.3 & \cite{fiore_sn_2021} \\
SN 2018bym & 58370.307 & MMT & Blue Channel & 0.15 & 109.0 & -1.2 & this work \\
SN 2018bym & 58425.217 & Gemini-North & GMOS-N & 0.31 & 152.1 & -0.2 & this work \\
\enddata
\tablecomments{The phase with respect to the peak of the bump is given in units of the bump FWHM. See Appendix for details on the new spectroscopy presented in this work.}
\end{deluxetable*}
\vspace{-24pt}

\subsection{Spectra\label{sec:spec}}
Both the scenarios mentioned above may have spectroscopic signatures. Circumstellar interaction is usually accompanied by narrow emission lines from the shocked CSM. Indeed, \cite{yan_hydrogen-poor_2017} have observed hydrogen emission in three Type~I SLSN spectra at approximately right phases (${\sim}$100 days after peak).\footnote{The two SLSNe discussed by \cite{yan_hydrogen-poor_2017} are included in our sample. SN~2016wi (a.k.a.\ iPTF15esb) shows hydrogen emission about a month after its light-curve bump. iPTF16bad shows hydrogen emission without a light-curve bump.} Changes in the ejecta ionization structure would also manifest as changing spectroscopic features during the bump, which have not previously been observed.

Although there has not been a concerted effort to obtain spectroscopic observations coeval with a light-curve bump, we investigate whether any of the SNe in our gold sample have public spectra taken in the duration of their bumps. Figure~\ref{fig:bumpspec} shows three such SNe. In each case, we plot the latest spectrum taken before the light-curve bump (outside the FWHM duration) and a spectrum taken near the peak of the bump. These spectra are logged in Table~\ref{tab:bumpspec}. In none of these cases do we observe a drastic change in the spectrum that might indicate the cause of the light curve excess.

This seems to indicate that the bumps have smooth spectra, rather than being caused by increases or decreases in discrete lines. One implication of this might be that the continuum emission and line-forming regions are quite distinct, i.e., the change in opacity happens in a dense shell where the photosphere lives, but there is no change in ionization state of the low-density gas forming the lines (see, e.g., \citealt{chen_evolution_2017}). Alternatively, the regions forming the continuum and the lines could be distinct due to extreme asymmetry, although that may not be consistent with the low levels of polarization observed in SLSNe (see Section~\ref{sec:csm}). It will be critical to increase this sample size with additional spectra of SLSNe at these late phases in order to draw a robust conclusion.

\section{Summary and Conclusions\label{sec:conclude}}
We have presented the first systematic analysis of a sample of late-phase ``bumps'' in SLSN light curves. Our primary conclusion is that a large fraction of SLSNe show such bumps, meaning, although their light curves can broadly be explained by the magnetar spin-down model, additional physics is required to reproduce their photometric behavior in detail.

By characterizing these bumps with an amplitude, temperature, phase, and duration, and then searching for correlations between these characteristics and the properties of the underlying magnetar model, we investigate two possible origins of the excess: interaction between the SN ejecta and a shell of CSM, and variability associated with the central engine. In the CSM model, we can constrain the requisite shell to be ${\sim}0.1\ M_\odot$ lying at ${\sim}10^5\ R_\odot$ from the progenitor star, with a ${\sim}10\%$ thickness. On the other hand, if the excess is intrinsic to the magnetar-powered ejecta, it tends to occur at the same evolutionary phase, about 3.7 rise times after explosion, when the photospheric temperature is 6000--8000~K. We find that most bumps are bluer than their underlying magnetar model.

Ultimately, we do not find any evidence to favor one model over the other for the majority of SLSNe. This type of detailed analysis of light-curve behavior is severely limited by a lack of high-cadence, late-phase photometry on a rare class of SNe. Fortunately, upcoming wide--fast--deep surveys like the Legacy Survey of Space and Time at Vera C.\ Rubin Observatory \citep{ivezic_lsst:_2019} will provide high-quality light curves of a large sample of SLSNe (if we can identify them; \citealt{villar_superluminous_2018}). Survey strategies also affect the types of bumps that we can detect. For example, greater depth will enable bump detections at later phases, and higher cadence will allow detection of shorter-duration bumps. However, a deep understanding of this behavior likely also requires a concerted effort to obtain late-phase spectroscopy, as well as radio and X-ray observations, for a significant fraction of these discoveries.

In addition to a lack of data, our analysis is hindered by a lack of detailed theoretical models of the two scenarios under consideration. Although both ejecta--CSM and magnetar--ejecta interaction have been considered in the literature, analytical models are restricted to a handful of the simplest cases. Robust (perhaps numerical) modeling of each scenario for a large number of SLSNe could greatly improve on the phenomenological arguments presented here.

\clearpage
We thank Manos Chatzopoulos for helpful discussions regarding models of magnetar-powered SNe and circumstellar interaction. The Berger Time-domain Group at Harvard is supported in part by the NSF under grant AST-1714498 and by NASA under grant NNX15AE50G. G.H.\ thanks the LSSTC Data Science Fellowship Program, which is funded by LSSTC, NSF Cybertraining grant \#1829740, the Brinson Foundation, and the Moore Foundation; his participation in the program has benefited this work. B.D.M.\ acknowledges support from the NSF (grant no.\ AST-2002577). M.N.\ is supported by a Royal Astronomical Society Research Fellowship and by the European Research Council under the European Union's Horizon 2020 research and innovation program (grant agreement No.~948381). Las Cumbres Observatory observations were taken as part of the Global Supernova Project. Based on observations obtained at the international Gemini Observatory, a program of NSF's NOIRLab, which is managed by the Association of Universities for Research in Astronomy (AURA) under a cooperative agreement with the National Science Foundation on behalf of the Gemini Observatory partnership: the National Science Foundation (United States), National Research Council (Canada), Agencia Nacional de Investigaci\'{o}n y Desarrollo (Chile), Ministerio de Ciencia, Tecnolog\'{i}a e Innovaci\'{o}n (Argentina), Minist\'{e}rio da Ci\^{e}ncia, Tecnologia, Inova\c{c}\~{o}es e Comunica\c{c}\~{o}es (Brazil), and Korea Astronomy and Space Science Institute (Republic of Korea). The computations in this paper were run on the FASRC Cannon cluster supported by the FAS Division of Science Research Computing Group at Harvard University.

\facilities{ADS, FLWO:1.2m (KeplerCam), FLWO:1.5m (FAST), Gemini:Gillett (GMOS; GN-2018B-FT-111), LCOGT (Sinistro), McGraw-Hill (Templeton), MMT (Blue Channel), NED, OSC, TNS}

\defcitealias{astropycollaboration_astropy_2018}{Astropy Collaboration 2018}
\software{Astropy \citepalias{astropycollaboration_astropy_2018}, \texttt{lcogtgemini} \citep{mccully_cmccully/lcogtgemini:_2021}, \texttt{lcogtsnpipe} \citep{valenti_diversity_2016}, Light Curve Fitting \citep{hosseinzadeh_light_2020}, Matplotlib \citep{hunter_matplotlib:_2007}, MOSFiT \citep{guillochon_mosfit:_2018}, NumPy \citep{oliphant_guide_2006}, PyRAF \citep{sciencesoftwarebranchatstsci_pyraf:_2012}, \texttt{pymccorrelation} \citep{privon_hard_2020}, PyZOGY \citep{guevel_pyzogy_2017}}

\appendix
\begin{deluxetable*}{lcCCcc}
\tablecaption{New Photometry of SLSNe\label{tab:phot}}
\tablehead{\colhead{SN} & \colhead{MJD} & \colhead{Filter} & \colhead{Apparent Magnitude} & \colhead{Telescope} & \colhead{Instrument}}
\startdata
SN 2017egm & 57908.187 & g & 15.547 \pm 0.005 & Las Cumbres 1.0m & Sinistro \\
SN 2017egm & 57908.192 & r & 15.884 \pm 0.008 & Las Cumbres 1.0m & Sinistro \\
SN 2017egm & 57908.195 & i & 16.063 \pm 0.019 & Las Cumbres 1.0m & Sinistro \\
SN 2017egm & 57914.171 & i & 15.602 \pm 0.009 & Las Cumbres 1.0m & Sinistro \\
SN 2017egm & 57914.175 & r & 15.477 \pm 0.007 & Las Cumbres 1.0m & Sinistro \\
SN 2017egm & 57914.180 & g & 15.207 \pm 0.006 & Las Cumbres 1.0m & Sinistro \\
SN 2017egm & 57915.141 & g & 15.133 \pm 0.005 & Las Cumbres 1.0m & Sinistro \\
SN 2017egm & 57915.146 & i & 15.593 \pm 0.014 & Las Cumbres 1.0m & Sinistro \\
SN 2017egm & 57915.150 & r & 15.376 \pm 0.005 & Las Cumbres 1.0m & Sinistro \\
SN 2017egm & 57915.175 & V & 15.170 \pm 0.014 & Las Cumbres 1.0m & Sinistro \\
SN 2017egm & 57915.181 & B & 15.201 \pm 0.018 & Las Cumbres 1.0m & Sinistro \\
SN 2017egm & 57922.150 & g & 14.570 \pm 0.070 & F.~L.~Whipple 1.2m & KeplerCam \\
SN 2017egm & 57922.150 & r & 14.850 \pm 0.060 & F.~L.~Whipple 1.2m & KeplerCam \\
SN 2017egm & 57922.150 & i & 15.060 \pm 0.080 & F.~L.~Whipple 1.2m & KeplerCam \\
SN 2017egm & 57922.700 & i & 15.190 \pm 0.040 & McGraw-Hill 1.3m & Templeton \\
SN 2017egm & 57922.710 & g & 14.650 \pm 0.040 & McGraw-Hill 1.3m & Templeton \\
SN 2017egm & 57922.710 & r & 15.020 \pm 0.040 & McGraw-Hill 1.3m & Templeton \\
SN 2017egm & 57922.710 & z & 15.380 \pm 0.070 & McGraw-Hill 1.3m & Templeton \\
\enddata
\tablecomments{This table is available in its entirety in machine-readable form.}
\end{deluxetable*}

\vspace{-28pt}

\section*{Additional Observations and Data Reduction\label{sec:appendix}}
Aside from published and publicly available photometry, we obtained additional photometric observations of SNe~2017egm, 2018bym, 2018fcg, 2019neq, and 2019ujb. These data are listed in Table~\ref{tab:phot}, which is available in machine-readable form. The majority of observations were taken in the \textit{gri} bands using KeplerCam on the 1.2\,m telescope at Fred Lawrence Whipple Observatory \citep{szentgyorgyi_keplercam_2005}. We also obtained three epochs of \textit{BVgri} imaging of SN~2017egm using the Sinistro camera on Las Cumbres Observatory's 1\,m telescope at McDonald Observatory  \citep{brown_cumbres_2013}, one epoch of \textit{griz} imaging of SN~2017egm using the Templeton camera on the 1.3\,m McGraw-Hill Telescope at MDM Observatory, and one epoch of $r$-band imaging of SN~2018bym (from the acquisition image for our spectrum) using the Gemini Multi-object Spectrograph (GMOS) on the Gemini-North telescope \citep{hook_gemini-north_2004}.

Images from KeplerCam and Templeton were reduced using standard PyRAF routines \citep{sciencesoftwarebranchatstsci_pyraf:_2012}. For KeplerCam, we subtracted archival images from the Pan-STARRS1 $3\pi$ survey \citep{chambers_pan-starrs1_2016} using HOTPANTS \citep{becker_hotpants:_2015}. We then extracted point-spread function (PSF) photometry with the \texttt{daophot} package and calibrated it to the $3\pi$ catalog.

Images from Las Cumbres Observatory were preprocessed using BANZAI \citep{mccully_lcogt/banzai:_2018}. PSF photometry was extracted using \texttt{lcogtsnpipe} \citep{valenti_diversity_2016}, which is based on PyRAF. The \textit{BV} filters were calibrated to the AAVSO Photometric All-sky Survey (APASS; \citealt{henden_aavso_2009}) and \textit{gri} to Data Release 12 of the Sloan Digital Sky Survey \citep{alam_eleventh_2015}.

\defcitealias{planck_collaboration_planck_2016}{Planck Collaboration (2016)}
Throughout this work, \textit{BV} magnitudes are in the Vega system and \textit{griz} magnitudes are in the AB system. Distance moduli were calculated using cosmological parameters from the \citetalias{planck_collaboration_planck_2016}. Milky Way extinction corrections are derived using the dust maps of \cite{schlafly_measuring_2011} with a \cite{cardelli_relationship_1989} extinction law. A constant $K$-correction of $2.5\log(1+z)$ was used for all filters.

The two spectra of SN~2018bym plotted in Figure~\ref{fig:bumpspec} are also presented here for the first time. The first was obtained with the Blue Channel spectrograph on the MMT telescope \citep{angel_optical_1979} and the second with GMOS on Gemini-North. Both were observed at the parallactic angle and reduced via standard PyRAF tasks to subtract bias frames, apply flat fields, model and subtract the sky spectrum, extract the target spectrum, calibrate the wavelength to an arc lamp spectrum, and calibrate the flux to a standard star spectrum. Specifically, the Gemini reduction followed the procedure of \cite{mccully_cmccully/lcogtgemini:_2021}. These spectra are available on the Weizmann Interactive Supernova Data Repository \citep{yaron_wiserep_2012}.

\bibliography{main.bib}

\end{document}